\documentclass[aps,pre,preprint,groupedaddress,showpacs]{revtex4-1}
\usepackage{amssymb}

\usepackage[dvips]{graphicx}
\usepackage{textcomp}
\usepackage{amssymb}

\newcommand{\mc}{\multicolumn}

\begin{document}

\title{ 
\Large\bf Thermodynamic Casimir effect: Universality and Corrections to Scaling}

\author{Martin Hasenbusch}
\email[]{Martin.Hasenbusch@physik.hu-berlin.de}
\affiliation{
Institut f\"ur Physik, Humboldt-Universit\"at zu Berlin,
Newtonstr. 15, 12489 Berlin, Germany}

\date{\today}

\begin{abstract}
We study the thermodynamic Casimir force for films in the three-dimensional
Ising universality class with symmetry breaking boundary conditions.  We focus 
on the effect of corrections to scaling and probe numerically the universality
of our results. In particular we check the hypothesis that corrections are
well described by an effective thickness 
$L_{0,eff}=L_0+c (L_0+L_s)^{1-\omega} +L_s$, where $c$ and $L_s$ are system 
specific parameters and $\omega\approx 0.8$ is the exponent of the 
leading bulk correction.
We simulate the improved Blume-Capel model and the spin-1/2 Ising
model on the simple cubic lattice.
First we analyse the behaviour 
of various quantities at the critical point. Taking into account corrections
$\propto L_0^{-\omega}$ in the case of the Ising model, we find
good consistency of results obtained from these two different models. 
In particular  we get from the analysis of our data for the Ising model
for the difference of Casimir amplitudes
$\Delta_{+-}-\Delta_{++}=3.200(5)$, which nicely compares with 
$\Delta_{+-}-\Delta_{++}=3.208(5)$ obtained by studying the improved 
Blume-Capel model. Next we study the behaviour of the 
thermodynamic Casimir force for large values of the scaling variable 
$x=t [L_0/\xi_0]^{1/\nu}$. It can be obtained up to an overall 
amplitude by  expressing the partition function of the film in terms
of eigenvalues and eigenstates of the transfer matrix and boundary
states. Here we demonstrate how this amplitude can be computed with 
high accuracy. 
Finally we discuss our results for the scaling functions 
$\theta_{+-}$ and $\theta_{++}$ of the thermodynamic Casimir force for 
the whole range of the scaling 
variable. We conclude that our numerical results are in accordance with
universality. Corrections to scaling are well approximated by an 
effective thickness. 
 
\end{abstract}
\maketitle

\section{Introduction}
At  a second order phase transition various quantities like 
the correlation length $\xi$ or the specific heat $C_{bulk}$ diverge following 
power laws such as
\begin{equation}
\label{powerlaw}
 \xi \simeq \xi_{0,\pm} |t|^{-\nu} \;\;,
 \;\;\;\;  C_{bulk} \simeq A_{\pm} |t|^{-\alpha}
\end{equation}
where $t=(T-T_c)/T_c$ is the reduced temperature, $\nu$ and $\alpha$ are
the critical exponents of the correlation length and the specific heat,
respectively. The indices $\pm$ of the amplitudes $\xi_{0,\pm}$ and 
$A_{\pm}$ indicate the phase: $+$ for the high temperature phase and $-$ 
for the low temperature phase. Critical exponents such as $\nu$ and $\alpha$
and amplitude ratios  such as $\xi_{0,+}/\xi_{0,-}$ and $A_+/A_-$ are 
universal. This means that these quantities do not depend on the microscopic
details of the system but are exactly the same for all systems within 
a universality class. A universality class is characterized by the dimension
of the system, the range of the interaction and the symmetry properties of
the order parameter. For reviews
on critical phenomena see e.g. \cite{WiKo,Fisher74,Fisher98,PeVi02}.
Power laws such as eq.~(\ref{powerlaw}) are valid only asymptotically in the 
limit $t \rightarrow 0$. At finite reduced temperature corrections 
have to be taken into account \cite{wegner71,Wegner-76}
\begin{equation}
\label{wegnerexpansion}
  \xi = \xi_{0,\pm} |t|^{-\nu} \times \left(1 + a_{\pm} |t|^{\theta} + b t
   + c_{\pm} |t|^{2 \theta} + d_{\pm} |t|^{\theta'} + ... \right) \;.
\end{equation}
There are analytic and non-analytic (confluent) corrections. The non-analytic 
corrections are associated with non-trivial exponents $\theta = \nu \omega$, 
$\theta' = \nu \omega'$, ... . For the universality class of the 
three-dimensional Ising model with short ranged interactions 
one finds consistently $\omega \approx 0.8$ 
from field theoretic methods, the analysis of high temperature series 
expansions and  Monte Carlo simulations of lattice models \cite{PeVi02}.
Our recent estimate is $\omega =0.832(6)$ \cite{mycritical}. 
The estimate $\omega'=1.67(11)$ obtained by 
the scaling field method \cite{NewmanRiedel} still lacks confirmation by other
approaches. Furthermore we expect corrections caused by the breaking of
symmetries by the lattice. In the case of the simple cubic lattice that we 
consider here, 
these corrections are associated with $\omega'' \approx 2$  \cite{pisa97}.

The singular behaviour~(\ref{powerlaw}) requires that the thermodynamic 
limit is taken.  For finite systems, the behaviour of thermodynamic quantities 
is given by analytic functions  of the parameters of the system and its 
linear size $L_0$. Finite size scaling \cite{Barber} predicts that in the 
neighbourhood of the critical point, for sufficiently large $L_0$, 
this behaviour
is characterized by a universal function of certain combinations of
the parameters of the system and its linear size $L_0$. 
In the absence of an external field, a quantity $A(L_0,t)$ that is a 
function of the temperature and the linear size $L_0$ of the system behaves as
\begin{equation}
\label{finiteLaw}
 A(L_0,t)  \simeq L_0^y \; g(t [L_0/\xi_{0,+}]^{1/\nu}) 
\end{equation}
where the function $g(x)$  depends on the universality class of the bulk 
system and on the geometry of the finite system and $y =w/\nu$, where
$A(\infty,t) \propto |t|^{-w}$. Also finite size scaling is affected by
corrections to scaling  \cite{Barber}
\begin{equation}
\label{finiteLawC}
 A(L_0,t)  = L_0^y \; g(t [L_0/\xi_{0,+}]^{1/\nu}) \;
   \left[1 +b \; q(t [L_0/\xi_{0,+}]^{1/\nu}) \; L_0^{-\omega} + ... \right]
\end{equation}
where $q(x)$ is a universal function and $b$ depends on the details of 
the system.

Here we shall study films with symmetry-breaking boundary conditions. 
This choice is motivated by the fact that for classical binary liquid mixtures, 
typically the surfaces are more attractive for one of the two components of the 
mixture. 
In addition to the corrections discussed above, these boundary conditions give
rise to additional corrections, where the leading one is $\propto L_0^{-1}$ 
\cite{BinderS,Diehl86,Diehl97}, 
where $L_0$ is now the thickness of the film. In this work we focus on the 
thermodynamic Casimir effect \cite{FiGe78} in films. Due to the fact that 
in the neighbourhood of the critical point the
range of thermal fluctuations is restricted by the finite thickness of the 
film an effective force arises. The thermodynamic Casimir
force per area is given by
\begin{equation}
\label{defineFF}
 F_{Casimir} = -   \frac{ \partial \tilde f_{ex} }{ \partial L_0} \;\;,
\end{equation}
where $\tilde f_{ex} = \tilde f_{film} - L_0 \tilde f_{bulk}$
is the excess free energy per area of the film, where $\tilde f_{film}$ is
the free energy per area of the film and $\tilde f_{bulk}$ the free energy 
density of the bulk system.
The thermodynamic Casimir force per area follows the finite size scaling law
\begin{equation}
\label{FCffs}
 F_{Casimir} \simeq k_B T L_0^{-3} \; \theta(t [L_0/\xi_{0,+}]^{1/\nu}) \;\;,
\end{equation}
see e.g. refs. \cite{Krech,Dantchev1}.  For a discussion of  non-universal
contributions due to long-ranged tails of the interaction, which is not 
the subject of the present paper, we refer the reader to ref.
 \cite{Dantchev2}.
After the seminal work \cite{FiGe78} it took about two decades until 
the thermodynamic Casimir effect could be demonstrated in experiments.
The data obtained for films of different thicknesses of $^4$He near the 
$\lambda$-transition are represented to a reasonable approximation
by a unique finite size scaling function \cite{GaCh99,GaScGaCh06}. 
Also experiments with liquid binary mixtures near the mixing-demixing
transition were performed, where either
films \cite{FuYaPe05} or the sphere-plate geometry \cite{Nature,GaMaHeNeHeBe09}
were  studied. Unfortunately, field theoretic methods do not allow to
compute the scaling function $\theta(x)$ for the full range of the 
scaling variable \cite{Diehl86,Diehl97}. Therefore it was an important
achievement that recently the thermodynamic Casimir force was computed by
Monte Carlo simulations of lattice models. Corresponding to
the experiments on $^4$He,  the XY model on the simple cubic lattice
was simulated \cite{Hucht,VaGaMaDi08}.  Also the Ising model on the 
simple cubic lattice that shares the universality class of the mixing-demixing 
transition of binary mixtures was studied \cite{VaGaMaDi07,VaGaMaDi08}. 
A reasonable match of the universal scaling functions $\theta$ obtained from  
experiments and the corresponding Monte Carlo simulations of lattice models
was found. For a recent review see \cite{Ga09}.

However it turned out that it is quite difficult to obtain precise results
for the universal scaling function $\theta$ from these Monte Carlo simulations.
For the thicknesses that can be reached, corrections to scaling are still 
significant. Fitting the data it is difficult to disentangle corrections 
$\propto L^{-\omega}$ and $\propto L^{-1}$. Furthermore the universal function
$q(x)$, eq.~(\ref{finiteLawC}), that governs the corrections 
$\propto L^{-\omega}$ is a priori unknown. The authors of 
\cite{Hucht,VaGaMaDi07,VaGaMaDi08} used ad hoc approximations
of $q(x)$ in the analysis of their data. Depending on the particular ansatz
that they used, the results of \cite{VaGaMaDi07,VaGaMaDi08} for the universal
scaling function vary by a large amount.

In order to alleviate this problem we \cite{myXY,mybreaking} 
studied improved models which are characterized by the 
fact that the amplitude of the leading bulk correction vanishes. Since the
parameter of the improved model is determined numerically, in practice a
residual amplitude remains, which is however at least by a factor of $30$ 
smaller than that of the Ising model and the XY model on the simple cubic 
lattice, respectively \cite{mycritical,ourXY}. Our results for the 
scaling functions of the thermodynamic Casimir force agree qualitatively
with those of refs. \cite{Hucht,VaGaMaDi07,VaGaMaDi08}. However the numerical
discrepancies are considerably larger than the errors that are quoted.
In particular, the results obtained very recently in \cite{VaMaDi11} from
simulations of the Ising model by
using the prefered ansatz of the authors, eqs.~(17,18) of \cite{VaMaDi11}, 
deviate  clearly from those of \cite{mybreaking}; See fig. 6 a of 
\cite{VaMaDi11}; and from that of \cite{mycrossing}; See fig. 6 b of 
\cite{VaMaDi11}. For a discussion of this fact by the authors 
of \cite{VaMaDi11}, see the text on page 041605-9 of  \cite{VaMaDi11}
starting about 20 lines below table II.

The aim of the present work is to reach a better  understanding of 
corrections to scaling. This means that we intend to determine the function
$q(x)$ of eq.~(\ref{finiteLawC}) for the thermodynamic Casimir force. 
Note that due to universality of the function $q(x)$ our results
might also be useful in the analysis of data obtained in experiments.
Also here we start with an ansatz for $q(x)$ which is motivated as follows.
The corrections $\propto L_0^{-1}$ caused by the boundaries 
can be expressed by a constant shift in the thickness of the film. In 
equations such as eq.~(\ref{finiteLaw}) the thickness $L_0$ is replaced by
\begin{equation}
\label{leff1}
 L_{0,eff} = L_0 + L_s  \;\;,
\end{equation}
where $L_s$ depends on the details of the system but not on the 
observable. Here we shall probe the hypothesis that in an analogue way 
corrections $\propto L_0^{-\omega}$  can be taken into account by
\begin{equation}
\label{leff2}
 L_{0,eff} = L_0  + c (L_0 +L_s)^{1-\omega} +  L_s \;\;.
\end{equation}
While renormalization group arguments suggest that eq.~(\ref{leff1}) is 
indeed exact, the generalization is at best a good approximation. It is
motivated by the fact that for the strongly symmetry breaking boundary
conditions studied here fluctuations are suppressed in the neighbourhood
of the boundaries. Hence the effect of corrections to scaling should
be the largest close to the boundaries.  Plugging eq.~(\ref{leff2}) 
into eq.~(\ref{finiteLaw}), ignoring the correction  $\propto L_0^{-1}$
due to the boundary, we get 
\begin{eqnarray}
 A(L_0,t)  &=& (L_0+c L_0^{1-\omega})^y \; 
    g(t [(L_0+c L_0^{1-\omega})/\xi_{0,+}]^{1/\nu}) \nonumber \\
          &=&
L_0^y \; g(x) \times \left (1+ c \left[y + \frac{x}{\nu} \frac{g'(x)}{g(x)}
\right] 
 L_0^{-\omega} + \mbox{O}( L_0^{-2 \omega}) \right)
\end{eqnarray}
where $x=t [L_0/\xi_{0,+}]^{1/\nu}$. Hence our hypothesis~(\ref{leff2}) 
results in 
\begin{equation}
\label{qansatz}
q(x)=y +  \frac{x}{\nu} \frac{g'(x)}{g(x)} \;.
\end{equation}

The outline of the paper is the following: In section \ref{themodel} we
define the models that we simulated and the observables that we 
measured. In section \ref{fssf} we briefly recall how finite size scaling 
theory applies to the free energy per area and the thermodynamic Casimir force
per area.
In section \ref{fssY} we study various quantities exactly at the 
critical point. Next, in section \ref{transfermatrix} we study the behaviour
of the  thermodynamic Casimir force for large values of the scaling 
variable $x$. To this end, we analyse the magnetisation profile near the 
boundary of the film and the correlation function of the bulk system.
In section \ref{thetafull} we discuss our results for the scaling 
functions $\theta_{++}$ and $\theta_{+-}$ in the full range of the 
scaling argument. Then we summarize and discuss our results. Finally
in the appendix we discuss various results obtained for the bulk of the 
spin-1/2 Ising model.

\section{Model}
\label{themodel}
We study the Blume-Capel model on the simple cubic lattice. 
It is defined by the reduced Hamiltonian 
\begin{equation}
\label{Isingaction}
H = -\beta \sum_{<xy>}  s_x s_y
  + D \sum_x s_x^2  \;\; ,
\end{equation}
where the spin might assume the values $s_x \in \{-1, 0, 1 \}$. 
$x=(x_0,x_1,x_2)$
denotes a site on the simple cubic lattice, where $x_i \in \{1,2,...,L_i\}$ 
and $<xy>$ denotes a pair of nearest neighbours on the lattice.
The inverse temperature is denoted by $\beta=1/k_B T$. The partition function 
is given by $Z = \sum_{\{s\}} \exp(- H)$, where the sum runs over all spin
configurations. The parameter $D$ controls the
density of vacancies $s_x=0$. In the limit $D \rightarrow - \infty$
vacancies are completely suppressed and hence the spin-1/2 Ising
model is recovered.

In  $d\ge 2$  dimensions the model undergoes a continuous phase transition
for $-\infty \le  D   < D_{tri} $ at a $\beta_c$ that depends on $D$.
For $D > D_{tri}$ the model undergoes a first order phase transition.
The authors of \cite{DeBl04} give for the three-dimensional simple cubic 
lattice $D_{tri}=2.0313(4)$.

Numerically, using Monte Carlo simulations it has been shown that there 
is a point $(D^*,\beta_c(D^*))$ 
on the line of second order phase transitions, where the amplitude
of leading corrections to scaling vanishes.  Our recent 
estimate is $D^*=0.656(20)$ \cite{mycritical}.  In \cite{mycritical} we
simulated the model at $D=0.655$ close to $\beta_c$ on lattices of a 
linear size up to $L=360$. From a standard finite size scaling analysis 
of phenomenological couplings like the Binder cumulant we find 
$\beta_c(0.655)=0.387721735(25)$. Furthermore the amplitude of leading 
corrections to scaling is at least by a factor of $30$ smaller than 
for the spin-1/2 Ising model. As discussed in the appendix \ref{appendixbc}
we  shall use $\beta_c=0.22165462(2)$ as estimate of the inverse critical
temperature of the spin-1/2 Ising model in the following.

In \cite{myamplitude} we 
simulated the Blume-Capel model at $D=0.655$  in the high 
temperature phase on lattices of the size $L^3$ with periodic boundary
conditions in all directions and $L \gtrapprox 10 \xi$ for 201 values
of $\beta$. For a few values of $\beta$ we performed new simulations 
that reduced the statistical error considerably. In particular for 
$\beta=0.3872$, which was our value closest to $\beta_c$,
we get $\xi_{2nd}(0.3872)=26.7013(15)$ for second moment correlation
length now.
Taking into account these new data we arrive at the slightly revised result
\begin{eqnarray}
\label{xi0}
\xi_{2nd,0,+} &=&  0.2283(1) - 1.8 \times (\nu-0.63002)
                        + 275 \times (\beta_c - 0.387721735) \;\; \nonumber \\
&&  \mbox{using} \;\; t = \beta_c - \beta \;\;
 \mbox{as definition of the reduced temperature} .
\end{eqnarray}
The analogue result for the  spin-1/2 Ising model is given in 
eq.~(\ref{amplitudexi}) in Appendix \ref{appendixamp}.

In the high temperature phase there is little difference between
$\xi_{2nd}$ and the exponential correlation length $\xi_{exp}$ which
is defined by the asymptotic decay of the two-point correlation function.
Following  \cite{pisaseries}:
\begin{equation}
\lim_{t\searrow 0} \frac{\xi_{exp}}{\xi_{2nd}} = 1.000200(3)  
\;\;
\end{equation}
for the thermodynamic limit of the three-dimensional system. 
Note that in the following $\xi_{0}$ always refers to $\xi_{2nd,0,+}$.

\subsection{Film geometry and boundary conditions}
In the present work we study the thermodynamic Casimir effect 
for systems with film 
geometry. In the ideal case this means that the system has a finite 
thickness $L_0$, while in the other two directions the thermodynamic 
limit $L_1, L_2 \rightarrow \infty$ is taken. In our  Monte Carlo 
simulations we shall study lattices with $L_0 \ll L_1, L_2$ and 
periodic boundary conditions in the $1$ and $2$ directions.  Throughout 
we shall simulated lattices with $L_1=L_2=L$. 

In the 0 direction we take symmetry breaking boundary conditions. 
A strong breaking of the symmetry  is achieved by fixing
the spins at the boundary to either
$-1$ or $1$. Here we shall put these fixed spins on the layers
at $x_0 = 0$ and at $x_0=L_0+1$. This means that $L_0$ gives the 
number of layers with fluctuating spins. In the following we shall 
consider the two choices:
\begin{itemize}
 \item $++$ boundary conditions:  $s_x=1$ for all $x$ with  $x_0=0$ or 
                                  $x_0=L_0+1$.
 \item $+-$ boundary conditions:   $s_x=1$ for all $x$ with  $x_0=0$
                                   and $s_x=-1$ for all $x$ with  $x_0=L_0+1$.
\end{itemize}
Note that these boundary conditions are physically relevant
for experiments with confined near-critical binary mixtures, since typically
a surface is more attractive to one of the components than to the 
other. In experiments, see e.g., \cite{FuYaPe05,Nature,GaMaHeNeHeBe09}, both possible 
situations can been realized: Both surfaces
prefer the same component or one surfaces prefers one component while the other
surfaces prefers the other component of the mixture. 

\subsection{Free energy, energy and specific heat}
For bulk systems we define the reduced free energy density as 
\begin{equation}
\label{fbulk}
f_{bulk} = - \frac{1}{L_0 L_1 L_2} \ln Z \;.
\end{equation}
This means that compared with the free energy density $\tilde f_{bulk}$,
a factor $k_B T$ is skipped.
Correspondingly we define the energy density as the derivative of
minus the reduced free energy density with respect to $\beta$
\begin{equation}
\label{Ebulk}
E_{bulk} = \frac{1}{L_0 L_1 L_2} \frac{\partial \ln Z}{\partial \beta}
   = \frac{1}{L_0 L_1 L_2}  
    \left \langle \sum_{<x,y>} s_x  s_y \right \rangle \; ,
\end{equation}
and the specific heat 
\begin{equation}
\label{Cbulk}
C_{bulk} = \frac{\partial E_{bulk}}{\partial \beta} = \frac{1}{L_0 L_1 L_2}
\left [
\left  \langle \left(  \sum_{<x,y>} s_x  s_y 
     \right)^2 \right \rangle  -
    \left \langle \sum_{<x,y>} s_x  s_y \right 
    \rangle^2 \right] \; .
\end{equation}
In the case of films we consider the reduced free energy per area 
\begin{equation}
\label{fdef1}
f = - \frac{1}{L_1 L_2} \ln Z \;.
\end{equation}
and the energy per area 
\begin{equation}
\label{Edef1}
E = \frac{1}{L_1 L_2} \frac{\partial \ln Z}{\partial \beta}
  = \frac{1}{L_1 L_2}  \left \langle \sum_{<x,y>} s_x  s_y \right \rangle \; .
\end{equation}

\subsection{The magnetization profile of films}
\label{mag}
The film is
invariant under translations in the 1 and 2 direction of the lattice.
Therefore the magnetization only depends on $x_0$ and we can average
over $x_1$ and $x_2$:
\begin{equation}
 m(x_0) \;=\; \frac{1}{L^2}  \sum_{x_1,x_2}  \langle s_x  \rangle \;\;.
\end{equation}
Since the film is symmetric for $++$ boundary conditions and
anti-symmetric for $+-$ boundary conditions under reflections at the middle
of the film, $m(x_0) = m(L_0-x_0+1)$ for $++$ boundary conditions and
$m(x_0) = - m(L_0-x_0+1)$ for $+-$ boundary conditions.

\subsection{The correlation length}
\label{xisection}
The exponential correlation length $\xi$ of the bulk system is defined by 
the decay of the slice-slice  correlation function
\begin{equation}
 G(r) \simeq c \exp(-r/\xi) 
\end{equation}
for large distances $r$. The slice-slice correlation function is defined as
\begin{equation}
 G(r) = \langle S(x_0) S(x_0+r)  \rangle 
\end{equation}
where 
\begin{equation}
  S(x_0) = \frac{1}{\sqrt{L_1 L_2}} \sum_{x_1,x_2} 
\left(s_{x_0,x_1,x_2} -  \langle m \rangle  \right)
\end{equation}
where $\langle m \rangle$ is the bulk magnetisation that vanishes in the 
high temperature phase, for a vanishing external field.

For a detailed discussion of the second moment correlation length defined 
for films see section III C of \cite{mybreaking}.

\subsection{Monte Carlo algorithms}
\label{MonteCarlo}
In the case of the Ising model we simulated films with $L_0 \le 68$
using a local 
Metropolis algorithm and a multispin coding implementation. We used the 
same program, up to small modifications to implement the boundary conditions,
as discussed in section 3 of ref. \cite{HaPi93}. 
On one core of an Intel(tm) Xeon(tm) E5520 CPU
running at  2.27 GHz the program achieves $1.9  \times 10^9$ spin updates 
per second. This is about 100 times faster than on the fastest workstation
that was available to us in 1993. Most simulations were performed on 
Quad-Core AMD Opteron(tm) 2378 CPUs running at 2.4 GHz. Here
the program achieves $1.4  \times 10^9$ spin updates per second on one core.
In relation with section \ref{transfermatrix} we simulated films
with $++$ boundary conditions with $L_0 > 68$. These were simulated by 
using a special version of the cluster algorithm as discussed ref. 
\cite{mybreaking}.
In the case of the Blume-Capel model we simulated the films using  
the same algorithms as discussed in section V of ref. \cite{mybreaking}.

Mostly we simulated lattices with periodic boundary conditions in 
all directions 
with the single-cluster algorithm \cite{Wolff} in the case
of the Ising model and a hybrid \cite{BrTa89} of the local heat-bath 
and the single-cluster algorithm in the case of the Blume-Capel model.

In all our simulations we used the
Mersenne twister algorithm \cite{twister} as pseudo-random number generator.
In total our simulations took the equivalent of about 50 years of CPU
time on a single core of a 
Quad-Core AMD Opteron(tm) Processor 2378 running at 2.4 GHz.

\section{Finite size scaling and corrections to scaling}
\label{fssf}
The reduced excess free energy per area of a film is given by
\begin{equation}
 f_{ex}(L_0,\beta) = f(L_0,\beta) - L_0 f_{bulk}(\beta) \;.
\end{equation}
In the reduced excess free energy the analytic bulk contribution cancels.
Therefore it can be written as 
\begin{equation}
 f_{ex}(L_0,\beta) = f_{ex,s}(L_0,\beta) + 2 f_r(\beta) 
\end{equation}
where $f_{ex,s}$ is the singular part and $f_r$ is an analytic 
contribution due to the boundaries. In the absence of an external field, 
this contribution is the same for a boundary where all spins are 
fixed to $+1$ 
and one where all spins are fixed to $-1$.  The free energy of a system
is conserved under renormalization group transformations. Therefore
the singular part of the reduced excess free energy behaves as 
\begin{equation}
 f_{ex,s}(L_0,\beta) = L_0^{-2} H(t [L_0/\xi_0]^{y_t}, b L_0^{y_1},...)
\end{equation}
where $y_t=1/\nu$ and $y_1=-\omega$ are the thermal and the leading 
irrelevant renormalization group exponent, respectively. Expanding the 
universal scaling function $H(x,u,...)$ in $u$ around $u=0$ we arrive at
\begin{equation}
 f_{ex,s}(L_0,\beta) = L_0^{-2} h(x) 
  \; \left[1+ b \; p(x) L_0^{-\omega} + ... \right]
\end{equation}
where $x=t [L_0/\xi_0]^{1/\nu}$ and the leading correction is characterized 
by the universal function $p(x)$.
Taking minus the derivative with respect to $L_0$ we get the thermodynamic
Casimir force
\begin{equation}
\frac{1}{k_b T} F_{Casimir} = L_0^{-3} \theta(x) \;
\left[1 + b \; \left(p(x) +  
\frac{h(x)}{\theta(x)} \left[\omega p(x) 
   - \frac{x}{\nu} p'(x) \right] \right)  L_0^{-\omega} + ...
 \right]
\end{equation}
where 
\begin{equation}
\theta(x)=2  h(x) - \frac{x}{\nu}  h'(x)  \;.
\end{equation}
Note that at the critical point $\theta(0)=2  h(0)$. In the literature 
$h(0)$ is called Casimir amplitude and is denoted by $\Delta$. Also
note that 
\begin{equation}
\label{energycasimir}
\theta'(0) = \left[2 -\frac{1}{\nu}\right] h'(0)  \;.
\end{equation}

Taking minus the derivative with respect to  $\beta$ we get
\begin{equation}
\label{energyFSS}
 E_{ex}(L_0,\beta) = L_0^{-2} [L_0/\xi_0]^{1/\nu}  h'(x)
       \left[1 + b \left(p(x) + \frac{h(x) p'(x)}{h'(x)} \right) 
       L_0^{-\omega} + ... \right ]
                                 -2 f_r'(\beta)  \;.
\end{equation}

\section{Finite size scaling at the critical point}
\label{fssY}

First we study finite size scaling at the critical point, 
i.e. $x = t [L_0/\xi_0]^{1/\nu} = 0$. To this end we analyse data for 
the free energy difference between films with $+-$ and $++$ boundary
conditions, the energy density and 
the magnetisation profile for both types of boundary conditions.
Finally we also consider the second moment correlation length for
$+-$ boundary conditions. 

For a given quantity at a given value of $x$ it is a trivial recast 
to express corrections to scaling in the form~(\ref{leff2}). The 
non-trivial question that we investigate here is whether leading 
corrections in different quantities can be expressed by the same or
at least similar effective thicknesses $L_{0,eff}$. 

In the ansaetze below we shall use in addition to eq.~(\ref{leff2})
\begin{equation}
\label{leff3}
 L_{0,eff} = L_0 + L_s + c (L_0 + L_s)^{-\omega} + d (L_0 + L_s)^{-\epsilon}
\end{equation}
in order to probe for the effect of subleading corrections. As discussed
in the intoduction, there are infinitely many subleading corrections 
starting with 
$\epsilon = 2 \omega \approx \omega'$, $1 + \omega$ and $\omega'' \approx 2$.
Given the accuracy of our data, it is only possible to put one subleading 
correction in the ansatz. In the following we shall take either 
$\epsilon=1.664$ or $\epsilon=2$.  Fitting with ansaetze that only
approximate the behaviour of the data one has to be aware of systematical
errors. In the literature it is often implicitly assumed that an acceptable
$\chi^2/$d.o.f. means that such systematical errors are small and of a similar
size or even smaller than the statistical errors of the fit parameters.
However this is definitely not the case. The severity of the problem 
depends of course on the type of the approximation and the range 
of the data that are available.  Below we shall see that the
differences between results of fits with eq.~(\ref{leff2}) 
and ones with eq.~(\ref{leff3}) are  e.g. five times larger than the 
statistical error. The error that we quote for final results is chosen such
that both the results of fits with eq.~(\ref{leff2}) and eq.~(\ref{leff3}) are
covered.

\subsection{The difference of free energies per area between
$+-$ and $++$ boundary conditions}
\label{Dffff}
First we studied the difference 
\begin{equation}
D_{f,+-,++}(L_0,\beta) = f_{+-}(L_0,\beta) - f_{++}(L_0,\beta) \;\;,
\end{equation}
where $f_{+-}$ and $f_{++}$ are the reduced free energies for
$+-$ and $++$ boundary conditions, respectively.
In this difference the surface and the bulk contributions exactly 
cancel and therefore at the critical point
\begin{equation}
\label{Dfatbetac}
 D_{f,+-,++}(L_0,\beta_c) \simeq (\Delta_{+-} - \Delta_{++}) \; L_0^{-2} \;\;,
\end{equation}
where $\Delta_{+-}$ and $\Delta_{++}$ are the Casimir amplitudes 
for $+-$ and $++$  boundary conditions, respectively.
Similar to the case of periodic and anti-periodic boundary conditions
\cite{MH93A,MH93B}, the ratio $Z_{+-}/Z_{++}$ of partition functions
can be directly computed 
by using the cluster algorithm.  To this end one determines for $++$
boundary conditions the fraction of cluster decompositions where
the two boundaries do not belong to the same cluster. These cluster 
decompositions would allow to update to $+-$ boundary conditions.
Since for $+-$ boundary 
conditions the update to $++$ boundary conditions is always allowed, the 
fraction discussed above is an estimate of $Z_{+-}/Z_{++}$. 

Unfortunately, at the critical point, for  $L \gg L_0$, the ratio 
$Z_{+-}/Z_{++}$ is far too small to allow for an efficient sampling.
Therefore we simulated in the high temperature phase at $\beta=\beta_0$
such that $L_0/\xi(\beta_0) \approx 6$, where $\xi$ is the bulk correlation
length. Here, for $L=4 L_0$, which we  used in our simulations, the value 
of $Z_{+-}/Z_{++}$ is a few percent.
In order to get $f_{+-} - f_{++}$ at larger values of $\beta$, 
in particular at the critical point, we performed an integration of 
energy differences:
\begin{equation}
\label{integ2}
D_{f,+-,++}(L_0,\beta) = D_{f,+-,++}(L_0,\beta_0) - 
\int_{\beta_0}^{\beta} \mbox{d} \tilde \beta \; D_{E,+-,++}(L_0, \tilde \beta)
\;\;,
\end{equation}
where $D_{E,+-,++} = E_{+-}-E_{++}$. We performed this integration 
numerically, using the trapezoidal rule. To this end, we  used at least
36 values of $\beta$ between $\beta_0$ and $\beta_c$ as nodes. For a detailed
discussion of the corresponding Monte Carlo simulations see section 
\ref{thetafull} below. In most cases we used the same data as discussed
in section \ref{thetafull}. Only for the Ising model at the thicknesses
$L_0=24$ and $48$ and the Blume-Capel model at the thickness $L_0=68$
we performed additional simulations.
For an analytic integrand, the estimate obtained by using the trapezoidal 
rule behaves as $I(h)= I(0) +  a h^2 + O(h^4)$,
where $I(0)$ is the integral to be computed and $h$ is the step-size.
We estimated the  systematic error by computing $I(2 h)$, i.e.
performing the integration~(\ref{integ2}) with half of the available data 
points. The systematic error is then estimated by $\epsilon=(I(2 h) - I(h))/3$.
It turned out that the systematic error $\epsilon$ is considerably
larger than the 
rather small statistical error. Therefore, we extrapolated our result as
$I(0) = I(h)-[I(2 h) - I(h)]/3 + O(h^4)$.  In the case of the Blume-Capel
model and $L_0=34$,  where we simulated at 116 values of $\beta$ 
between $\beta_0$ and $\beta_c$ we checked the efficiency of the 
extrapolation by computing $I(h)$, $I(2 h)$ and $I(4 h)$. We found agreement
between $I(h)-(I(2 h) - I(h))/3$ and $I(2 h)-(I(4 h) - I(2 h))/3$ within 
the statistical error.
In table \ref{DFtable} we  summarized our numerical results for the 
critical point.

\begin{table}
\caption{\sl \label{DFtable}  We give the difference $D_{f,+-,+-}$ of the 
reduced free energies per area between $+-$ and $++$ boundary conditions at
our estimates of the inverse critical temperature,
i.e. $\beta=0.22165462$ for the Ising model and $\beta=0.387721735$ for the 
Blume-Capel model at $D=0.655$.
}
\begin{center}
\begin{tabular}{ccl}
\hline
$L_0$ & Model &\mc{1}{c}{$ D_{f,+-,+-}$}  \\
\hline
14  & I & 0.01069953(37)  \\
15  & I & 0.00953606(25)  \\ 
16  & I & 0.00855417(15)  \\ 
17  & I & 0.00771682(12)  \\ 
24  & I & 0.00423239(15)  \\
32  & I & 0.002522796(50) \\ 
34  & I & 0.002258418(55) \\
48  & I & 0.00119288(10) \\
64  & I & 0.000693495(64) \\ 
68  & I & 0.000617863(63) \\  
\hline
16 & BC & 0.00999910(67) \\  
17 & BC & 0.00897065(65) \\
32 & BC & 0.00279016(11) \\
34 & BC & 0.00248788(11) \\
68 & BC & 0.00065641(11) \\
\hline
\end{tabular}
\end{center}
\end{table}

We fitted the data obtained for the Ising model with the ansaetze
\begin{equation}
\label{diffansatz1}
 D_{f,+-,++}  = \Delta \; [L_0 + L_s + c (L_0+L_s)^{1-\omega} ]^{-2}
\end{equation}
and 
\begin{equation}
\label{diffansatz2}
 D_{f,+-,++}  =\Delta \; 
[L_0 + L_s + c (L_0+L_s)^{1-\omega}+d (L_0+L_s)^{1-\epsilon}]^{-2}  \;\;,
\end{equation}
where we set either $\epsilon=1.664$ or $\epsilon=2$. 

Fitting with the ansatz~(\ref{diffansatz1}), setting $\omega=0.832$ we get for 
$L_{0,min}=16$ the result
$\Delta = 3.1987(9)$, $c=1.429(12)$, $L_s=1.043(12)$ and $\chi^2/$d.o.f.$=1.20$.
Note that all data with $L_0 \ge L_{0,min}$ are taken into account in the fit.
Instead, taking $\omega=0.826$  we get 
$\Delta=3.1995(9)$, $c=1.367(11)$, $L_s=1.100(15)$ and $\chi^2/$d.o.f.$=1.20$.
This means that the estimate of $\Delta$ depends little on the value of 
$\omega$, while $c$ and $L_s$ are quite sensitive to it.
We  redid these fits for $D_{f,+-,++}$ evaluated at $\beta=0.2216546$.
The results change only by little. 

Next we fitted all data, i.e. $L_{0,min}=14$, 
with the ansatz~(\ref{diffansatz2}).  We get, fixing $\omega=0.832$
and $\epsilon=1.664$ the results  $\Delta=3.2025(21)$, $c=1.57(6)$, 
$L_s=0.78(11)$, $d=0.35(14)$ and $\chi^2/$d.o.f.$=1.47$. 
Instead, for $\epsilon=2$ we get 
$\Delta=3.2016(19)$, $c=1.52(4)$, $L_s=0.88(7)$, $d=1.52(4)$ and 
$\chi^2/$d.o.f.$=1.45$.  We see that by adding a subleading correction the 
estimate of $\Delta$ changes little, while the results for $c$ and $L_s$ 
are considerably shifted. Note that the estimates of $c$ and $L_s$ are 
highly 
anti-correlated. The resulting $L_{0,eff}$, eq.~(\ref{leff2}), for the 
thicknesses analysed here, depend much less on the ansatz that is used.
Taking all fits discussed above into account we conclude
\begin{equation}
\label{IsingDelta}
\Delta_{+-}-\Delta_{++}= 3.200(5)  \;\;.
\end{equation}

Next we fitted our data for the Blume-Capel model with the ansaetze
\begin{equation}
\label{murks1}
 D_{f,+-,++}  = \Delta \; [L_0 + L_s]^{-2}
\end{equation}
and
\begin{equation}
\label{murks2}
 D_{f,+-,++}  =\Delta \;
[L_0 + L_s +d (L_0+L_s)^{-1}]^{-2}  \;\;.
\end{equation}
Fitting all data with the ansatz~(\ref{murks1}) we
get $\Delta=3.20901(25)$, $L_s=1.9140(11)$ and $\chi^2/$d.o.f. $=1.12$. 
Fitting all data  with the ansatz~(\ref{murks2}) we get
$D=3.2071(5)$, $L_s=1.898(4)$, $d=0.20(6)$ and  $\chi^2/$d.o.f. $=0.72$, 
instead.
We redid these fits for $D_{f,+-,++}$ evaluated at $\beta=0.38772176$ in 
order to estimate the error due to the uncertainty of $\beta_c$.  
Finally, in order to check for the possible effect of residual corrections 
to scaling $\propto L_0^{-\omega}$, we fitted our data with the 
ansaetze~(\ref{diffansatz1},\ref{diffansatz2}), where we  fixed the
amplitude 
of the leading correction to $c=1.5/30$. Note that in ref. \cite{mycritical}
we found that the amplitudes of the leading correction are at least suppressed 
by the factor $1/30$ in the Blume-Capel model at $D=0.655$ compared with 
the spin-1/2 Ising model.

Taking these fits into account we arrive at 
\begin{equation}
\label{BCDelta}
\Delta_{+-}-\Delta_{++}=3.208(5) 
\end{equation}
which is consistent with the estimate~(\ref{IsingDelta}) obtained above.
Furthermore these results are fully consistent with
$\Delta_{+-}-\Delta_{++}= [\theta_{+-}(0)-\theta_{++}(0)]/2= 
[5.613(20)+0.820(15)]/2=3.217(18)$
obtained in section VI C of ref. \cite{mybreaking}. Our result is slightly 
larger than  $\Delta_{+-}-\Delta_{++}=2.71(2)-[-0.376(29)] =3.09(5)$ 
which the authors obtained by fitting their data for the thermodynamic
Casimir force per area with ansatz~(26) of ref. \cite{VaGaMaDi08}.
In \cite{VaMaDi11} the authors used different ansaetze.
Eqs.~(17,18,19) coincide at the critical point with our 
ansatz~(\ref{leff1}). The authors argue  that corrections 
$\propto L_0^{-\omega}$ are effectively taken into account by the
$\propto L_0^{-1}$ correction that is present in the ansatz.
In figure 6 a of \cite{VaMaDi11} we see that their strong symmetry 
breaking results, i.e. $\tilde h_1=-100$ and $\tilde h_1=100$ clearly
deviate from ours \cite{mybreaking}. To understand this discrepancy 
we fitted our data for the Ising model with the ansatz~(\ref{murks2}).
Fitting all our data  we get $\Delta=3.1467(4)$, $L_s=3.480(4)$,
$d=-5.83(5)$, and $\chi^2/$d.o.f.$=76.35$. Fitting only the data with 
$L_0 \le 34$ and assuming a statistical error that is 3 times larger 
than the one that we acctually achieved we get $\Delta=3.136(2)$, 
$L_s=3.39(2)$, $d=-4.7(2)$ and $\chi^2/$d.o.f.$=1.03$.  While 
$\chi^2/$d.o.f.$\approx 1$, this is completely incompatible with our  
final result~(\ref{IsingDelta}), 
which substantiates our statements above on fitting with appoximate ansaetze.

Finally note that 
our results for $L_s$ of the Blume-Capel model at $D=0.655$
are fully consistent with $L_s=1.9(1)$
\cite{mybreaking}, $L_s=2 l_{ex} = 1.92(4)$  and $L_s=1.90(5)$  
\cite{mycrossing}. In section \ref{thetafull} below, we shall assume
$L_s=1.91(5)$. 

\subsection{Simulations at the critical point} 
In order to compute the energy per area and the magnetisation profile 
at the critical point of the 
Ising model, we  performed high statistics simulations at 
$\beta=0.2216546$, which was our estimate of $\beta_c$ when we started 
the simulations. In order to obtain the observables at $\beta=0.22165462$,
we computed the 
derivate of the observables with respect to $\beta$ from finite 
differences.  In table \ref{stat} we summarize the lattice sizes and the 
statistics of our first set of simulations. 
\begin{table}
\caption{\sl \label{stat} 
Number of measurements (stat) in our simulations of the Ising model at
$\beta=0.2216546$. For each measurement 16 sweeps with the Metropolis
algorithm were performed. In these simulations $L = 6 L_0$ and
$L = 10 L_0$ for $++$ and  $+-$ boundary conditions, respectively.
}
\begin{center}
\begin{tabular}{rrr}
\hline
  \mc{1}{c}{$L_0$} &  \mc{1}{c}{stat $++$} & \mc{1}{c}{stat $+-$} \\
\hline
6  &  $64.0 \times 10^8 $ & $ 64.0 \times 10^7 $\\
7  &  $57.2 \times 10^8 $ & $ 64.0 \times 10^7 $\\ 
8  &  $45.3 \times 10^8 $ & $ 64.0 \times 10^7 $ \\  
9  &  $47.9 \times 10^8 $ & $ 64.0 \times 10^7 $\\   
10 &  $39.4 \times 10^8 $ & $ 51.5 \times 10^7 $\\
11 &  $31.4 \times 10^8 $ & $ 46.1 \times 10^7 $\\   
12 &  $24.0 \times 10^8 $ & $ 44.8  \times 10^7$ \\
13 &  $15.1 \times 10^8 $ & $ 37.9  \times 10^7$ \\
14 &  $15.5 \times 10^8 $ & $ 32.4 \times 10^7 $\\ 
15 &  $15.3 \times 10^8 $ & $ 27.8  \times 10^7$ \\  
16 &  $14.2 \times 10^8 $ & $ 25.6 \times 10^7 $\\   
17 &  $10.4 \times 10^8 $ & $ 21.5  \times 10^7 $ \\
18 &  $10.9 \times 10^8 $ & $ 19.5  \times 10^7 $\\
19 &  $11.8 \times 10^8 $ & $ 18.9 \times 10^7 $\\
20 &  $10.4 \times 10^8 $ & $ 18.8  \times 10^7 $\\
22 &  $10.4 \times 10^8 $ & $ 28.7  \times 10^7 $\\
24 &  $67.1 \times 10^7 $ & $ 20.4  \times 10^7 $\\
26 &  $64.3 \times 10^7 $ & $ 22.7  \times 10^7 $\\
28 &  $62.0 \times 10^7 $ & $ 25.8  \times 10^7 $\\
32 &  $62.9 \times 10^7 $ & $ 22.5  \times 10^7 $ \\ 
36 &  $31.5 \times 10^7 $ & $ 22.7   \times 10^7 $\\
48 &  $14.4 \times 10^7 $ & $ 2.9  \times 10^7 $\\
64 &  $9.9  \times 10^7 $ & $ 2.7 \times 10^7 $\\
\hline
\end{tabular}
\end{center}
\end{table}
In a second set of simulations
with $+-$ boundary conditions we measured the second moment 
correlation length in addition. We  simulated lattices of the 
thicknesses $L_0=24$, $32$, $48$, $64$ and $96$. The number of measurements
is $51.2 \times 10^7$, $51.9 \times 10^7$, $49.1 \times 10^7$, 
$34.6 \times 10^7$, and $7.8  \times 10^7$, respectively.
Also here we performed 16 sweeps with the Metropolis algorithm for 
each measurement.
For this second set of simulations $L=4 L_0$.  For  $L_0=6$ we  
simulated $L=12$, $14$, $16$, $18$, $20$, $24$, $36$, $48$ and $60$
performing $6.4 \times 10^8$ measurements throughout. From the 
analysis of these runs we conclude that for $+-$ boundary conditions, 
at the critical point $L=4 L_0$ is fully sufficient to keep deviations
from the $L \rightarrow \infty$ limit at a negligible level.  In our 
simulations we wrote averages over 64000 measurements on disc to 
keep the amount of data tractable.  In order to estimate autocorrelation 
times we did a few additional simulation, where every measurement was
stored. For example we performed $10^5$ measurements for $+-$ boundary
conditions, $L_0=96$ and $L=384$. From this run we got the integrated
autocorrelation times $\tau_{int}=3.3(2)$, $15.2(1.0)$ and $28.(3.)$ in 
units of measurements for the energy per area, the magnetic susceptibility
and the magnetisation in the middle of the film. The autocorrelation times
of a local algorithm grow like $\tau \propto L_0^{z}$ at the critical point, 
where $z \approx 2$.  Therefore, despite the efficient multispin coding 
implementation of the Metropolis algorithm, the cluster algorithm should 
become more efficient starting from a certain thickness $L_0$. Since 
$\tau_{int}$ enters into the statistical error this thickness depends
to some extend on the observable one is interested in.
For lack of human time, we did not systematically investigate these
questions. 

\subsection{The energy per area} 
Taking eq.~(\ref{energyFSS}) at $x=0$ and ignoring corrections to scaling 
we arrive at 
\begin{equation}
\label{Ebc}
 E_{ex}(L_0,\beta_c) = B + a L_0^{-2+1/\nu}
\end{equation}
where $B=2 f_r(\beta_c)$ and $a=\xi_0^{-1/\nu} h'(0)$.    

In order to compute the excess energy, we used the estimate of 
$E_{bulk}(\beta_c)$, eq.~(\ref{Enonsingular}), obtained in 
appendix \ref{appendixbc}. 
Replacing $L_0$ by $L_{0,eff}$
in eq.~(\ref{Ebc}) we arrive at the ansaetze
\begin{equation}
\label{Efit1}
E_{ex}(L_0,\beta_c) = B + 
a [L_0 + L_s + c (L_0 + L_s)^{1-\omega}]^{-2+1/\nu}
\end{equation}
and
\begin{equation}
\label{Efit2}
E_{ex}(L_0,\beta_c) = B  + 
a [L_0+L_s+c (L_0 + L_s)^{1-\omega}+d (L_0 + L_s)^{1-\epsilon}]^{-2+1/\nu}
\end{equation}
where we set either $\epsilon=1.664$ or $\epsilon=2$. In our fits, 
$B$, $a$, $c$, $L_s$ and $d$ are free parameters. We fixed 
$\nu=0.63002$ and $\omega=0.832$.

First we analysed our data for $+-$ boundary conditions.
Fitting with the ansatz~(\ref{Efit1}) we get an acceptable $\chi^2$/d.o.f.
starting from $L_{0,min}=18$. For $L_{0,min}=20$ we get $B=7.8010(7)$, 
$a=-15.455(7)$, $c=1.472(35)$, $L_s=1.413(44)$ and $\chi^2$/d.o.f.$=0.62$. 
Using the ansatz~(\ref{Efit2}) we get an acceptable $\chi^2$/d.o.f.
already for $L_{0,min}=7$ both for $\epsilon=1.664$ and $\epsilon=2$. 
For example for $L_{0,min}=8$ and $\epsilon=1.664$ we get 
$B=7.80405(23)$, $a=-15.4946(19)$, $c=2.028(8)$, $L_s=0.476(10)$, 
$d=-0.27(4)$ and $\chi^2$/d.o.f.$=0.93$. Instead for $L_{0,min}=8$ and 
$\epsilon=2$ we get $B=7.80279(25)$, $a=-15.4790(20)$, $c=1.794(9)$, 
$L_s=0.903(11)$, $d=0.13(4)$ and $\chi^2$/d.o.f.$=1.13$. We see that the 
results depend strongly on the ansatz that is used. This holds in particular 
for the estimates of $c$ and $L_s$. 
We redid the fits using shifted values of the 
input parameters to estimate the error of our results due to the 
uncertainty of these  parameters. 
Taking into account the results of all these fits we arrive at $B=7.803(5)$
and
\begin{equation}
a_{I,+-}=-15.48(5) - 130 \times (\nu-0.63002)
\end{equation}
where for $a_{+-}$ 
we give the dependence on the value of $\nu$ explicitly.
The error induced by the uncertainty of the other input parameters is 
included into the number given in $()$.

For $++$ boundary conditions fitting with the ansatz~(\ref{Efit1}) 
gives acceptable values of $\chi^2$/d.o.f. already for $L_{0,min}=7$. 
For example for $L_{0,min}=8$ we get $B=7.80168(22)$, $a=-10.2105(16)$, 
$c=1.462(7)$, $L_s=1.16(7)$ and $\chi^2$/d.o.f.$=1.23$.  Instead fitting
with the ansatz~(\ref{Efit2}) we get for $\epsilon=1.664$ and 
$L_{0,min}=8$ the result $B=7.8054(7)$, $a=-10.248(6)$, $c=2.02(4)$, 
$L_s=0.27(5)$, $d=2.3(4)$ and  $\chi^2$/d.o.f.$=1.02$. Fixing $\epsilon=2$
we get results that lie between those of the two fits discussed before.
Also in the case of  $++$ boundary conditions  we redid the fits
with shifted values of the input parameters.
As our final result we quote $B=7.804(5)$   and 
\begin{equation}
 a_{I,++} = -10.23(5)   - 70  \times (\nu-0.63002)  \;\;.
\end{equation}  
Note that the results obtained for $B$ with $+-$ and $++$ boundary 
conditions agree as theoretically expected.

Assuming $\nu=0.63002$ we get 
$h_{+-}'(0) = -15.48(5) \times 0.1962(1)^{1/0.63002}=-1.167(5)$
and $h_{++}'(0) = -10.23(5) \times 0.1962(1)^{1/0.63002}=-0.771(5)$ 
from the analysis of the Ising model.
In ref. \cite{mybreaking} we found for the Blume-Capel model at 
$D=0.655$ the results  $a_{BC,++}=-8.04(1)$ and  $a_{BC,+-}=-12.18(3)$. 
Hence 
$h_{+-}'(0) = -12.18(3) \times 0.2283(1)^{1/0.63002}= 1.168(4)$ and 
$h_{++}'(0) = -8.04(1) \times 0.2283(1)^{1/0.63002}= -0.771(2)$. We see
that the results obtained for the universal quantities $h_{+-}'(0)$ and 
$h_{++}'(0)$ are in perfect agreement. Using eq.~(\ref{energycasimir}) we
arrive at
\begin{equation}
 \theta_{+-}'(0) =-0.482(2)  \;\;,\;\;\;\;  \theta_{++}'(0) = -0.318(2)
\end{equation}
taking into account the results obtained from both models.

Finally we analysed the difference $D_{E,+-,++}$ at the critical point.
The advantage of this quantity is that the bulk energy and the surface 
contributions exactly cancel. We fitted our data with the ansaetze
\begin{equation}
\label{DEfit1}
E_{ex}(L_0,\beta_c) = 
D_a [L_0 + L_s + c (L_0 + L_s)^{1-\omega}]^{-2+1/\nu}
\end{equation}
and
\begin{equation}
\label{DEfit2}
E_{ex}(L_0,\beta_c) = 
D_a [L_0+L_s+c (L_0 + L_s)^{1-\omega}+d (L_0 + L_s)^{1-\epsilon}]^{-2+1/\nu}\;.
\end{equation}
Fitting with the ansatz~(\ref{DEfit1}) we get an acceptable $\chi^2$/d.o.f. 
only for rather large values of $L_{0,min}$. For example for $L_{0,min}=26$
we get $D_a=-5.2548(19)$, $c=1.80(10)$, $L_s=1.49(14)$ and  
$\chi^2$/d.o.f.$=1.26$. Fitting with  the ansatz~(\ref{DEfit2}) and 
$\epsilon=1.644$ we get for  $L_{0,min}=12$ the results
 $D_a=-5.2570(8)$, $c=2.15(4)$, $L_s=0.81(5)$, $d=-3.56(18)$ and 
$\chi^2$/d.o.f.$=1.04$. Instead for $\epsilon=2$ we get 
$D_a=-5.2548(7)$, $c=1.92(3)$, $L_s=1.25(4)$ and $d=-3.71(16)$ and 
$\chi^2$/d.o.f.$=1.09$. Also here, we redid the fits with shifted 
values of the input parameters. We arrive at the final result
\begin{equation}
 a_{I,+-} - a_{I,++} = -5.256(4)  - 75  \times (\nu-0.63002)  \;\;.
\end{equation}

\subsection{The magnetisation profile}
For simplicity, we shall not study the complete magnetisation profile,
but we shall restrict ourselfs on the magnetisation in the middle of 
the film and the slope of the magnetisation in the middle of the film 
for $++$ and $+-$ boundary conditions, respectively. 

Let us first discuss the case of $++$ boundary conditions. The magnetisation
in the middle of the film at the critical point behaves as 
\begin{equation}
 m_{mid} = C_{m} L_0^{-\beta/\nu}  \;.
\end{equation}
The amplitude $C_m$ is not universal, but one can construct universal 
amplitude ratios that combine $C_m$ with the amplitude of the bulk 
correlation length and the bulk magnetisation or the magnetic 
susceptibility.
Here we only intend to compare our result for $C_{m,I}$ for the Ising 
model with $C_{m,BC}$ obtained previously for the Blume-Capel model at 
$D=0.655$ \cite{mybreaking}. To this end it is sufficient to determine 
the relative normalization of the magnetisation between these two models.
To this end we compare the magnetic susceptibility of systems with 
the extension $L_0=L_1=L_2$ and periodic boundary conditions in all three 
directions that we computed in relation with ref. \cite{mycritical}.
In particular we fitted the data for the magnetic susceptibility
at $Z_a/Z_p= 0.5425$ with the ansatz
\begin{equation}
 \bar{\chi}= C_{\chi} L^{2-\eta} \times (1 + c L^{-\omega} ) + b
\end{equation}
where we fixed $\eta=0.03627(10)$ and $\omega=0.832(6)$. 
We arrive at
\begin{equation}
\label{fieldratio}
\sqrt{\frac{C_{\chi,I}}{C_{\chi,BC}}} =  1.2811(2)
\end{equation}
where statistical and systematical errors as well as the uncertainty of 
$\eta$ and $\omega$ are taken into account.

In order to define the magnetisation in the middle of the film for 
even values of the thickness $L_0$ we quadratically extrapolated the 
magnetisations of the slice that is next to the middle and 
the one that is next to next. We  fitted these data with the ans\"atze
\begin{equation}
\label{mag1}
m_{mid} = C_{m} (L_0 + L_s + c (L_0 + L_s)^{1-\omega})^{-\beta/\nu}
\end{equation}
and
\begin{equation}
\label{mag2}
m_{mid} = C_{m} (L_0 + L_s + c (L_0 + L_s)^{1-\omega}
                   +d (L_0 + L_s)^{1-\epsilon})^{-\beta/\nu}  \;\;.
\end{equation}
where we fixed $\beta/\nu=(1 + \eta)=0.5018135$, $\omega=0.832$ and 
$\epsilon=1.664$ or $\epsilon=2$. In the following we only take
into account data for even values of $L_0$.
Using ansatz~(\ref{mag1}) we get for $L_{0,min}=24$ the 
results $C_m=1.71799(18)$, 
$c=1.63(2)$, $L_s=1.31(3)$ and $\chi^2/$d.o.f.$=0.21$.
Using ansatz~(\ref{mag2})
we get with $\epsilon=2$ an acceptable $\chi^2/$d.o.f. already for 
$L_{0,min}=6$. For $L_{0,min}=8$ we get the results $C_m=1.71880(7)$,
$c=1.844(8)$, $L_s=0.922(11)$, $d=1.289(14)$ and $\chi^2/$d.o.f.$=0.78$. 
For $\epsilon=1.664$ and $L_{0,min}=10$ we get $C_m=1.71929(11)$, $c=2.016(17)$,
$L_s=0.563(29)$, $d=1.172(27)$ and $\chi^2/$d.o.f.$=0.59$. 

We redid these fits using shifted values of $\beta_c$, $\eta$ and 
$\omega$. As final results we quote
\begin{equation}
C_{m,I} = 1.7187(10) + 4.8 \times (\eta-0.03627)
\end{equation}
where we give explicitly the dependence of our result on the value 
of $\eta$.

In ref. \cite{mybreaking} we analysed $m_{mid}$ for the Blume-Capel
model at $D=0.655$ for thicknesses up to $L_0=32$. Later \cite{mycrossing} 
we added data for $L_0=48$, $64$ and $96$. Taking into account
also these data we arrive at
\begin{equation}
C_{m,BC} = 1.3421(8)  + 2.8 \times (\eta-0.03627)  
\end{equation}
We get
\begin{equation}
\frac{C_{m,I}}{C_{m,BC}}= 1.2806(16) 
\end{equation}
which is fully consistent with eq.~(\ref{fieldratio}).

In the case of $+-$ boundary conditions, we consider the slope of the 
magnetisation profile in the middle of the film. It scales as
\begin{equation}
 S_{mid} = C_s L_0^{-1-\beta/\nu} \;.
\end{equation}
We fitted our data for the Ising model with the ansaetze
\begin{equation}
\label{Smag1}
S_{mid} = C_s (L_0 + L_s + c (L_0 + L_{s})^{1-\omega})^{-1-\beta/\nu}
\end{equation}
and
\begin{equation}
\label{Smag2}
S_{mid} = C_s (L_0 + L_s + c (L_0 + L_{s})^{1-\omega}
  +  d (L_0 + L_{s})^{1-\epsilon} )^{-1-\beta/\nu}
\end{equation}
where we fixed $\eta=0.03627$ and $\omega=0.832$ and $\epsilon=1.664$
or $\epsilon=2$. Also here we fitted only the data for even 
values of $L_0$.
Fitting with the ansatz~(\ref{Smag1}) we find small values of 
$\chi^2/$d.o.f. 
already for $L_{0,min}=8$. For $L_{0,min}=10$ we get $C_{s,I}=7.2013(4)$, 
$c=1.4603(25)$, $L_s=0.7023(31)$ and $\chi^2/$d.o.f.$=0.39$. Fitting with 
the ansatz~(\ref{Smag2}) we find that the parameter $d$ vanishes within 
the error bars. Taking into account the error due to the uncertainty 
of the input parameters $\omega$ and $\eta$ we arrive at the 
\begin{equation} 
 C_{s,I} = 7.201(3) + 19  \times (\eta-0.03627) \;\;.
\end{equation}
Fitting data obtained in relation with ref. \cite{mybreaking} for  the 
Blume-Capel model we get 
\begin{equation}
C_{s,BC} = 5.625(10) + 10  \times (\eta-0.03627) \;\;.    
\end{equation}
We get
\begin{equation}
\frac{C_{s,I}}{C_{s,BC}}= 1.280(3)
\end{equation}
which is fully consistent with eq.~(\ref{fieldratio}).

\subsection{The correlation length}
Finally we discuss the second moment 
correlation length of films with $+-$ boundary conditions at the critical point.
Our numerical results are summarized in table  \ref{corranti}. Since  here we
generated less data than for the quantities discussed above we abstain 
from fitting the data for the correlation length.  In ref. \cite{mybreaking}
we found  $\xi_{2nd} = 0.2115(8) (L_0 + L_s)$.  Based on this result we 
define 
\begin{equation}
L_{0,eff} = \xi_{2nd} /0.2115(8) \;.
\end{equation}
In the third column of table \ref{corranti} we quote $L_{0,eff}-L_0$.
In $[]$ we give the error
due to the uncertainty of the amplitude of the correlation length 
of the film.
For comparison we give analogous results derived from the difference of 
free energies $D_{f,+-,++}$, the difference of energies $D_{E,+-,++}$, 
the magnetisation in the middle of the film for $++$ boundary conditions 
and the slope of the magnetisation in the middle of the film for 
$+-$ boundary conditions.

\begin{table}
\caption{\sl \label{corranti}   In the second column we give 
the second moment correlation length obtained from
simulations of lattices with $L=4 L_0 $ for $+-$ boundary 
conditions at the critical point of the Ising model. 
 In the third column we give $L_{ex}=L_{0,eff}-L_0$. For the 
definition of $L_{0,eff}$ see the text. In the fourth, fifth, sixth, 
and seventh column we give $L_{ex}=L_{0,eff}-L_0$ derived from
$D_{f,+-,++}$, $D_{E,+-,++}$, $m_{mid}$, and $S_{mid}$, respectively.
}
\begin{center}
\begin{tabular}{cclcccc}
\hline
 $L_0$ &$\xi_{2nd}$& $L_{ex}$, $\xi_{2nd}$ 
 &$L_{ex}$, $D_{f,+-,++}$& $L_{ex}$, $D_{E,+-,++}$&$L_{ex}$, $m_{mid}$&
 $L_{ex}$, $S_{mid}$ \\
\hline
 24 &  5.6881(24) & 2.89[10] & 3.51[2] & 4.61[10]& 4.14[3] & 3.20[1]\\ 
 32 &  7.4025(42) & 3.00[13] & 3.64[3] & 4.76[12]& 4.27[4] & 3.32[1]\\
 48 &  10.807(10) & 3.10[19] & 3.83[4] & 4.99[16]& 4.49[6] & 3.51[1]\\
 64 &  14.204(20) & 3.16[25] & 3.97[5] & 5.16[20]& 4.65[8] & 3.64[2]\\
 96 &  20.99(10)  & 3.2[4]   &   -     &   -     &   -     & 3.85[3]\\
\hline
\end{tabular}
\end{center}
\end{table}
We see that the values of $L_{0,eff}-L_0$ computed from different 
observables are of 
a similar size. However the differences are considerably larger than the 
sum of the errors. Therefore it is quite clear that $L_{0,eff}-L_0$
is not exactly the same for all quantities. 

\section{Thermodynamic Casimir force and the transfer matrix}
\label{transfermatrix}
First let us briefly recall the discussion given in section IV
of ref. \cite{mybreaking}. 
The partition function of a system with fixed boundary conditions 
can be expressed in terms of the eigenvalues of the transfer matrix 
and the overlap of the eigenvectors with the boundary states.
Let us consider a lattice of the size $L_0 \times L^2$, where 
$L$ is large compared with the bulk correlation length but still 
finite. 
We consider the transfer matrix $T$ that acts on vectors that are 
build on the configurations living on $L^2$ slices. We denote the 
eigenvalues of $T$ by $\lambda_{\alpha}$ and the corresponding 
eigenvector by $|\alpha \rangle $, where $\alpha=0,1,2,...,\alpha_{max}$. 
The eigenvalues are ordered such that $\lambda_{\alpha} \ge \lambda_{\beta}$
for $\alpha < \beta$. In particular $\lambda_{0}$ is the largest  eigenvalue.
The partition function of the system with fixed boundaries is given by
\begin{equation}
\label{transfer}
 Z_{b_1,b_2} = \sum_{\alpha} \lambda_{\alpha}^l \; \langle b_1 | \alpha \rangle
                                                \langle b_2 | \alpha \rangle
\;\;,
\end{equation}
where $l=L_0+1$ for our definition of the thickness $L_0$. 
The boundary states $b_{1,2}$ are either $+$  or $-$ here.
It follows that  
\begin{eqnarray}
\frac{L^2}{k_B T} F_{Casimir} &=& \frac{\partial}{\partial l} 
 \left[ \ln  Z_{b_1,b_2} - l \ln \lambda_0 \right] 
   =
\frac{\sum_{\alpha} \ln(\lambda_{\alpha}/\lambda_0)\;
(\lambda_{\alpha}/\lambda_0)^l \;
 \langle b_1 | \alpha \rangle \langle b_2 | \alpha \rangle}
{\sum_{\alpha} (\lambda_{\alpha}/\lambda_0)^l 
 \langle b_1 | \alpha \rangle \langle b_2 | \alpha \rangle} \nonumber \\
&=&
- \frac{\sum_{\alpha} m_{\alpha}  \exp(-m_{\alpha} l) \;
  \langle b_1 | \alpha \rangle \langle b_2 | \alpha \rangle}
  {\sum_{\alpha} \exp(-m_{\alpha} l)  \;
 \langle b_1 | \alpha \rangle \langle b_2 | \alpha \rangle} \;\;,
\end{eqnarray}
where 
$1/\xi_{\alpha} = m_{\alpha} = - \ln(\lambda_{\alpha}/\lambda_0)$ are
inverse correlation lengths.
In the high temperature phase for $\xi_1=\xi \ll L_0$ the
force is dominated by the contribution from $\alpha=1$. Hence
\begin{equation}
\tilde \theta(m l) \approx \frac{l^3}{k_B T} F_{Casimir} \approx
- m^3 l^3 \exp(-m l) \; \frac{1}{m^2 L^2} 
 \frac{\langle b_1 | 1 \rangle \langle b_2 | 1 \rangle}
      {\langle b_1 | 0 \rangle \langle b_2 | 0 \rangle} \;\;.
\end{equation}
The finite size scaling behaviour of the thermodynamic Casimir force implies
that 
\begin{equation}
C_b = \frac{1}{m L} \frac{\langle b | 1 \rangle}
                          {\langle b | 0 \rangle}
\end{equation}
has a finite scaling limit.  The state $|0 \rangle$ is symmetric 
under the global transformation $s_x \rightarrow -s_x$ for all $x$ in 
a slice, while  $|1 \rangle$ is 
anti-symmetric and therefore $C=C_+ = - C_-$. Hence
\begin{equation} 
\label{XZX}
\tilde \theta_{++}(m l) = - \tilde \theta_{+-}(m l) = 
- C^2 \; m^3 l^3 \exp(-m l)
\end{equation}
for sufficiently large values of $m l$. Since 
$x = t [l/\xi_0]^{1/\nu} \simeq (m l)^{1/\nu}$ it follows
\begin{equation}
\label{scalinghigh}
\theta_{++}(x) = - \theta_{+-}(x) \simeq - C^2  x^{3 \nu} \exp(-x^{\nu})
\end{equation} 
for sufficiently large values of $x$. 

\subsection{$C$ and the magnetisation profile}
In the following we shall discuss how the overlap amplitude $C^2$ can 
be computed from the magnetisation profile of a semi-infinite system with 
$+$ boundary conditions and the correlation function of slice magnetisations.
In terms of the transfer matrix, the magnetisation at position 
$x_0$ in a film of thickness $L_0$ is given by
\begin{equation}
\label{magT}
\langle M(x_0) \rangle = \left \langle \sum_{x_1,x_2} s_{x_0,x_1,x_2} \right \rangle = 
                 \frac{ \sum_{\alpha,\beta} 
                 \lambda_{\alpha}^{x_0} \lambda_{\beta}^{l-x_0}
                 \langle b_1 | \alpha \rangle 
                 \langle \alpha |\hat M| \beta \rangle 
                 \langle \beta | b_2 \rangle  }
                 {\sum_{\alpha}
                 \lambda_{\alpha}^{l}  
                 \langle  b_1 | \alpha \rangle
                 \langle \alpha | b_2 \rangle}  \;\;.
\end{equation}
In the basis of slice configurations, $\hat M$ is a diagonal matrix, 
where the elements give the magnetisation of the corresponding configuration. 
For $l \gg \xi$ and $\xi_2 \ll x_0 \ll l$ eq.~(\ref{magT}) reduces to
\begin{eqnarray}
\label{trapro}
 \langle M(x_0) \rangle &=& \frac{ 
                 \lambda_{1}^{x_0} \lambda_{0}^{l-x_0}
                 \langle b_1 | 1 \rangle
                 \langle 1 |\hat M| 0 \rangle
                 \langle 0 | b_2 \rangle  }
                 {
                 \lambda_{0}^{l}
                 \langle  b_1 | 0 \rangle
                 \langle 0 | b_2 \rangle} 
    = 
    \frac{\langle b_1 | 1 \rangle}{\langle  b_1 | 0 \rangle} 
       \langle 1 |\hat M| 0 \rangle  
       \left(\frac{\lambda_1}{\lambda_0} \right)^{x_0} \nonumber \\
    &=&
      m  L \; C_{b_1} \; \langle 1 |\hat M| 0 \rangle \; \exp(-m x_0) \;\;.
\end{eqnarray}
The quantity $O_M = \langle 1 |\hat M| 0 \rangle/L$ is finite in the limit
$L \rightarrow \infty$, since $\langle M(x_0) \rangle/L^2$ is finite in 
this limit. 

The slice-slice correlation function for a lattice of linear size $L_0$ and
periodic boundary conditions is given by
\begin{equation}
 G(r) =\frac{1}{L^2} \langle M(x_0) M(x_0+r) \rangle = 
  \frac{1}{L^2} \frac{\sum_{\alpha,\beta} 
   \langle \beta  |\hat M| \alpha \rangle \langle \alpha |\hat M| \beta \rangle 
    \lambda_{\alpha}^{r} \lambda_{\beta}^{L_0-r}}
   {\sum_{\alpha}  \lambda_{\alpha}^{L_0}}  \;\;.
\end{equation}
Since $\hat M$ is antisymmetric under $s_x \rightarrow -s_x$ for all $x$
in the slice, $\langle 0 |\hat M| 0 \rangle$ vanishes. 
For $\xi_2 \ll x_0 \ll L_0$ we get
\begin{equation}
\label{Gr1}
 G(r) = \frac{1}{L^2} \langle 0 |\hat M| 1 \rangle \langle 1 |\hat M| 0 \rangle
     \exp(-m r) = O_M^2 \exp(-m r) \;\;.
\end{equation}
Taking into account the periodicity  of the lattice we arrive at
\begin{equation}
\label{Gr2}
 G(r) = O_M^2
     \frac{\exp(-m r) + \exp(-m (L_0-r)) }{1 + \exp(-m L_0)} 
\end{equation}
which we shall use in our numerical analysis below.

\subsection{Numerical implementation}

In order to compute $G(r)$ we simulated lattices with $L_0=L_1=L_2=L$
and periodic boundary conditions. In the case of the Blume-Capel model we 
simulated the model by using a hybrid \cite{BrTa89} of the local heat-bath 
algorithm and the single-cluster algorithm \cite{Wolff}. In the case of the 
Ising model we only used the single-cluster algorithm. We measured the 
correlation function $G(r)$ by using its cluster-improved estimator. 
In order to keep deviations from the thermodynamic limit negligible we
chose $L > 10 \xi$ 
throughout. For a discussion of this point see section III or ref. 
\cite{myamplitude}.
In order to compute $\xi$ and $O_M^2$ from eq.~(\ref{Gr2}) we took the 
correlation function at the distance $r$ and $r+1$. For eq.~(\ref{Gr1})
one gets $\xi=1/\ln(G(r+1)/G(r))$ and $O_M^2=G(r) \exp(r/\xi)$. 
For eq.~(\ref{Gr2})
we solved the system of two equations numerically. We  computed the 
statistical errors of $\xi$ and $O_M^2$ and their covariance by using the 
Jackknife 
method. We checked which distance $r$ is needed to keep corrections 
due to eigenstates of the transfer matrix with $\alpha > 1$ negligible. 
As a result, we took $r \approx 2 \xi$ throughout.

In the case of the Blume-Capel model at $D=0.655$ we simulated at 11 values 
of $\beta$ between $\beta=0.34$ where $\xi=1.50420(13)$ and 
$\beta=0.3872$ where $\xi=26.7102(16)$.  For $\beta=0.3872$ we performed 
about $10^7$ update cycles. Each cycle consists of two sweeps of the 
local heat-bath
algorithm and $10^4$ single-cluster updates. Note that the average cluster size
at $\beta=0.3872$ is $1645.58(17)$, and hence the lattice of the size $270^3$ 
is covered on average $0.84$ times by these $10^4$ clusters.
The simulation at $\beta=0.3872$ took the equivalent of about 13 month of CPU-time on 
a single core of a  Quad-Core AMD Opteron(tm) Processor 2378 running at 2.4 GHz.
In the case of the Ising model, we simulated at 59 values of $\beta$ between 
$\beta=0.125$ where $\xi=0.667308(53)$ and $\beta=0.2208$ where 
$\xi=16.6711(12)$.

Next we analysed the magnetisation profile of films with $++$ boundary 
conditions. Also here we required that $L_i > 10 \xi$.  When possible, we  
used the results obtained from the simulations that we performed to compute 
the thermodynamic Casimir force. For values of $\beta$ where this is not the 
case, we performed extra simulations using the cluster algorithm.  
Taking $O_M^2$ and $\xi$ obtained above from the simulations of the lattices
with periodic boundary conditions as input one gets an estimate of 
$C(\xi)$ from eq.~(\ref{trapro}) for each distance $x_0$ from the boundary.
Throughout we took our final result from  $x_0 \approx 3 \xi$.  

\begin{figure}
\begin{center}
\includegraphics[width=14.5cm]{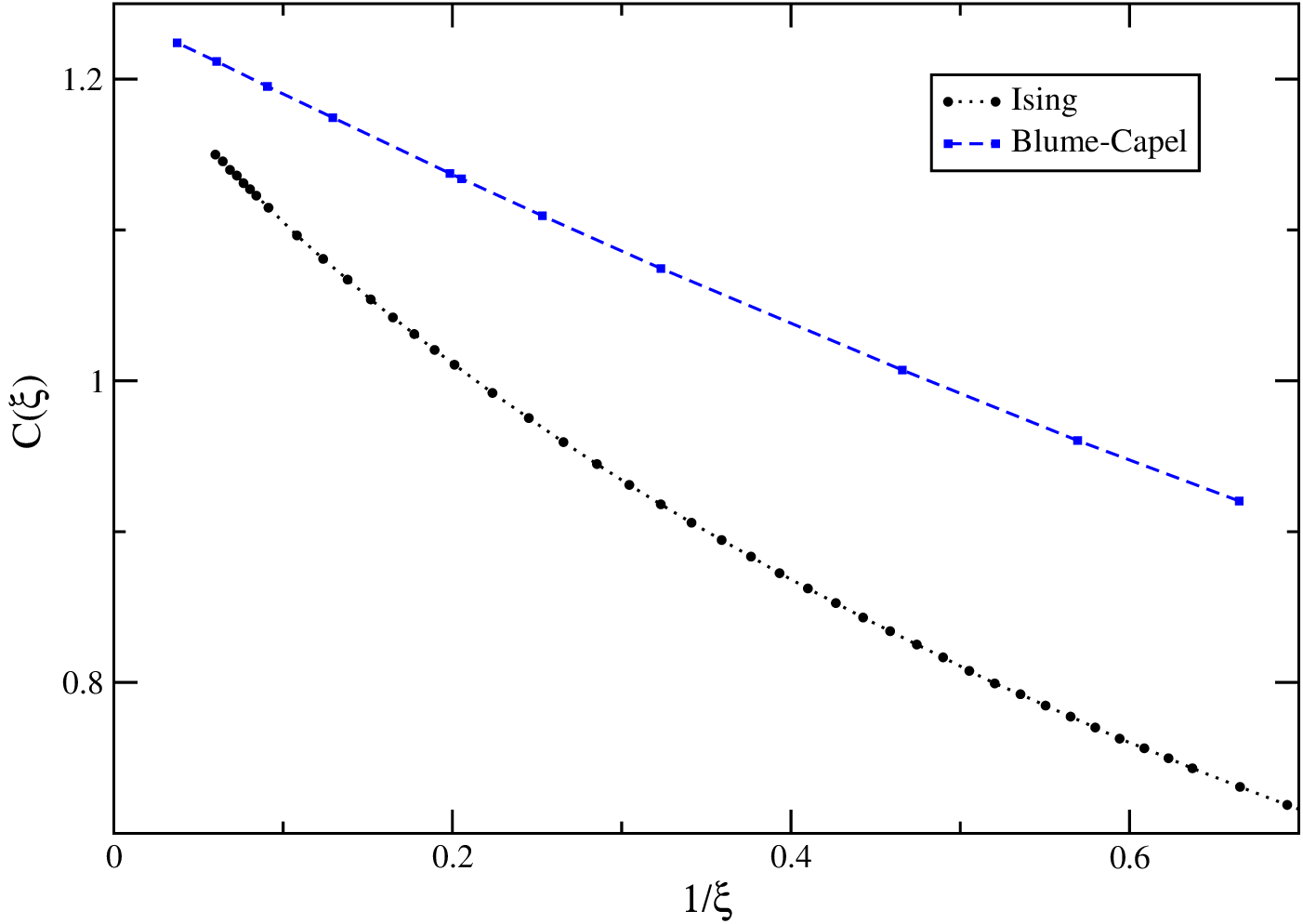}
\caption{\label{Cplot}
The amplitude $C(\xi)$ for the Ising model and the Blume-Capel model at 
$D=0.655$ as a function of $1/\xi$.
}
\end{center}
\end{figure}

In figure \ref{Cplot} we plot our results for $C(\xi)$ as a function of 
$m=1/\xi$ for the Ising model and the Blume-Capel model at $D=0.655$.
 Note that the error bars are much
smaller than the size of the symbols. For example for the Blume-Capel model at 
$\beta=0.3872$ we obtain $C(\xi) = 1.2241(4)$ and for the Ising model at 
$\beta=0.2208$ we get $C(\xi) = 1.1500(3)$.
The data for the  Blume-Capel model essentially fall on a straight line, 
confirming
that corrections $\propto \xi^{-\omega}$ are eliminated and those 
$\propto \xi^{-1}$ 
caused by the boundary dominate. In contrast, for the Ising model 
we see a clear
bending of the curve. It is conceivable that in the limit 
$\xi \rightarrow \infty$
the two curves converge to a unique value.

In order to substantiate these qualitative observations 
we fitted our data with the ansaetze
\begin{equation}
\label{Cansatz1}
 C(\xi) = C \exp(-c/\xi)
\end{equation}
and
\begin{equation}
\label{Cansatz2}
 C(\xi) = C \exp(-c/\xi) + a \xi^{-\epsilon}
\end{equation}
where $C$, $c$ and $a$ are the parameters of the fit. 
First we analysed our data for the Blume-Capel model.
Fitting with the ansatz~(\ref{Cansatz1}) we get $\chi^2$/d.o.f.$=0.67$, 
for fitting all data except the smallest value of $\beta$. The 
results for the parameters of the fit are $C=1.24568(21)$ and 
$c=0.4572(7)$. Next we fitted all data with the ansatz~(\ref{Cansatz2}).
Fixing $\epsilon=0.832$, we get $C=1.2462(5)$, $c=0.442(9)$,
$a=-0.017(11)$ and $\chi^2$/d.o.f.$=1.06$. For $\epsilon=2$ we get 
$C=1.24588(27)$, $c=0.4591(15)$, $a=0.0043(23)$ and 
$\chi^2$/d.o.f.$=0.64$.  As our final estimate we give
\begin{equation}
\label{Cfinal}
C = 1.2459(7)
\end{equation}
where the error-bar covers the results of the three fits given above.
The estimate $C^2=1.5(1)$ given in \cite{mybreaking} is consistent with, 
but much less precise than our present estimate $C^2 = 1.552(2)$
Note that the result $c \approx 0.46$ is fully consistent with 
$l_{ex}=0.96(2)$ obtained in \cite{mycrossing}. Note that for our 
definition of the thickness one expects $c = l_{ex}-1/2$. 

Next we fitted our data for the Ising model with the 
ansatz~(\ref{Cansatz2}) using $\epsilon=0.832$.  Fitting all data with 
$\beta \ge 0.202$ we get $C=1.24653(23)$, $a=-1.3750(29)$, 
$c=-0.479(2)$ and $\chi^2$/d.o.f.$=1.17$. Taking into  account smaller 
values of $\beta$, $\chi^2$/d.o.f. rapidly increases. We  redid the 
fit using $\epsilon=0.826$ and we also fitted with ansaetze that include
subleading corrections. Taking into account the results of these 
fits we arrive at $C=1.247(3)$, which is fully consistent with 
the result~(\ref{Cfinal}) that we obtained from the data for the 
Blume-Capel model.

We performed a similar study to determine the behaviour of the 
thermodynamic Casimir force for $++$ boundary conditions for 
$x \rightarrow -\infty$ in the low temperature phase. However here we can
not reach the same precision as above, since there is no efficient 
improved estimator for the correlation function in the low temperature
phase, and contributions due to subleading states of the 
transfer matix are more important than in the high temperature phase.
In the case of the Blume-Capel model we computed $\bar{C}$ for 16 
values of $\beta$ in the range from $\beta=0.39$ where $\xi=5.584(40)$ 
up to $\beta=0.405$ where  $\xi=1.5697(49)$. In the case of the Ising model
in the range from $\beta=0.223$ where $\xi=6.6028(20)$ up to 
$\beta=0.227$  where  $\xi= 2.7321(42)$.

Analysing the data for the Blume-Capel model, fixing $c=0.46(2)$ we arrive at 
$\bar{C}=0.428(10)$ and hence $\bar{C}^2=0.183(9)$ which 
is consistent with but more precise than $\bar{C}^2=0.20(5)$ given 
in \cite{mybreaking}.  Analysing the data for the Ising model, we get 
a consistent result.

\subsection{The correction function}
Plugging in $C^2(t) = C^2 (1+a_c t^{\theta})$ and 
$\xi=\xi_0 t^{-\nu} (1+a_{\xi} t^{\theta})$ into eq.~(\ref{XZX}) we get, e.g.
for $+-$ boundary conditions
\begin{eqnarray}
-\frac{\partial f_{ex}}{\partial L_0} 
 &=& L_0^{-3} C^2 \frac{L_0}{\xi_0} t^{\nu}
\exp\left(- \frac{L_0}{\xi_0} t^{\nu} \right) 
\times \left [1 + \left(a_c +  \left[\frac{L_0}{\xi_0} t^{\nu}-3\right] a_{\xi}
\right) t^{\theta}
+ \mbox{O}(t^{2 \theta}) \right] \nonumber \\
&=& L_0^{-3} \theta(x) \times \left [1 +  b \tilde q(x) L_0^{-\omega} 
         + \mbox{O}(L_0^{-2 \omega}) \right]
\end{eqnarray}
with
\begin{equation}
\label{qa}
 b \tilde q(x) = \xi_0^{\omega} (a_c + [x^{\nu} -3] a_{\xi}) x^{\theta} \;.
\end{equation}
which is not consistent with
\begin{equation}
\label{qb}
 b q(x) = - c x^{\nu}
\end{equation}
that one derives by plugging eq.~(\ref{scalinghigh}) into eq.~(\ref{qansatz}).
\begin{figure}
\begin{center}
\includegraphics[width=14.5cm]{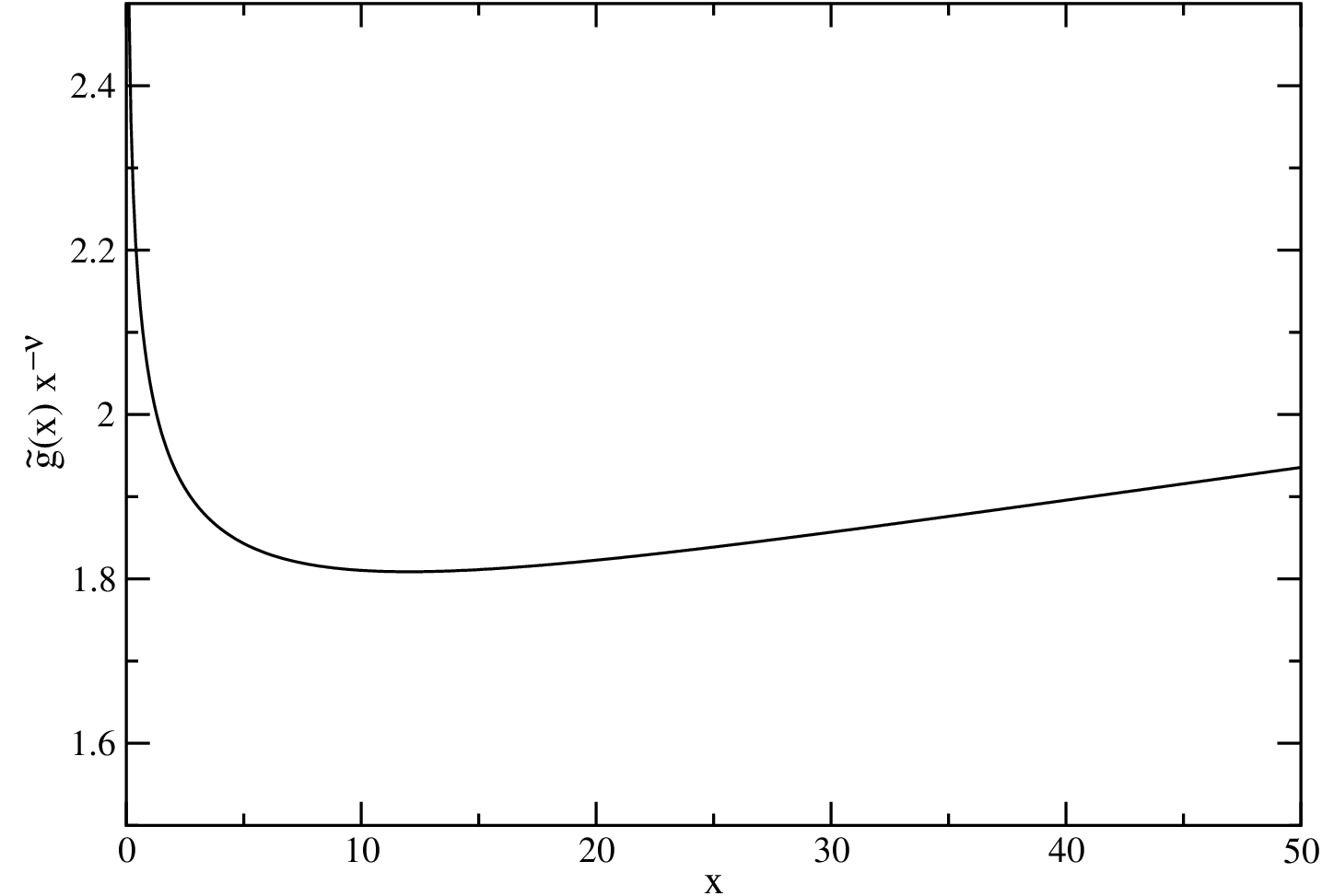}
\caption{\label{correct}
We plot $\tilde q(x) x^{-\nu}$ as a function of the scaling variable $x$
in the range that is relevant for our problem. For the definition of 
$\tilde q(x)$ and a discussion see the text.
}
\end{center}
\end{figure}
In figure \ref{correct} we plot $\tilde q(x) x^{-\nu}$ as a function of $x$. 
To this end, we take the numerical values $\xi_0=0.1962$,
$a_{\xi}=-0.32$, eq.~(\ref{correctionaxi}), and 
$a_C=2 \times 0.1962^{-0.832} \times (-1.375/1.247)= -8.55$.
It turns out that the curve is very flat in the range of $x$ we
are interested in. Also the value is rather close to the values of $c$
that we obtained from the analysis of data directly at the critical
point.

\section{The scaling function of the thermodynamic Casimir force for 
$++$ and $+-$ boundary conditions}
\label{thetafull}
We computed the thermodynamic Casimir force using the method discussed
by Hucht \cite{Hucht}. Starting from the energy per area we computed
\begin{equation}
\label{DeltaE}
\Delta E_{ex}(L_0,\beta)=[E(L_0+d/2,\beta)-E(L_0-d/2,\beta)]/d-E_{bulk}(\beta) \;\;.
\end{equation}
The value of the energy density of the bulk system $E_{bulk}(\beta)$
is obtained from an analysis of the high temperature series given in \cite{ArFu03}
and the low temperature series given in \cite{Vohwinkel} combined with Monte Carlo
simulations. For details see Appendix \ref{appendixene}.

In order to obtain $\Delta f_{ex}$ we numerically integrated
$\Delta E_{ex}$ using the trapezoidal rule:
\begin{equation}
\label{integration}
-\Delta f_{ex}(\beta_n) \approx 
-\Delta f_{ex}(\beta_0) +
\sum_{i=0}^{n-1} \frac{1}{2} (\beta_{i+1}-\beta_i)
   \left[\Delta E_{ex}(\beta_{i+1}) + \Delta E_{ex}(\beta_{i}) \right]
\end{equation}
where $\beta_{i}$ are the values of $\beta$ we simulated at. They
are ordered such that $\beta_{i+1} > \beta_i$ for all $i$.  In previous
work $\beta_0$ had been chosen such that $\Delta E_{ex}(\beta_{0}) \approx 0$  
and therefore also $\Delta f_{ex}(\beta_0) \approx 0$. 
Here, instead we chose a somewhat larger value of $\beta_0$ such that the 
approximation discussed in the previous section is still valid. In particular, 
we set 
\begin{equation}
\label{startingpoint}
\Delta f_{ex}(\beta_0) = \pm \frac{C^2(\beta_0)}{\xi^2(\beta_0)}
\frac{\exp[-(L_0 + 1+d/2)/\xi(\beta_0)] -\exp[-(L_0 + 1-d/2)/\xi(\beta_0)]}{d}
\end{equation}
where we have the $+$ sign for $++$ boundary conditions and the $-$
sign for $+-$ boundary conditions. By comparing results obtained with
different choices of $\beta_0$ we found that the 
approximation~(\ref{startingpoint}) is accurate at the level of our statistical
error up to $L_0/\xi(\beta_0) \gtrapprox 8$. To be on the safe side, 
we used $L_0/\xi(\beta_0) > 10$ in the following.

We simulated the Ising model with $++$ boundary conditions for the 
thicknesses $L_0=8$, $9$, $14$, $15$, $16$, $17$, $18$, $19$, $32$, $34$, $64$, 
and $68$. Using the resulting data we computed the thermodynamic Casimir 
force for
the thicknesses $L_0=8.5$ and $L_0=16.5$ using the difference $d=1$. In order
to check for the effect of using a finite difference to compute 
$\partial/\partial L_0$ we redid the calculation for $L_0=16.5$ using 
$d=3$ and $5$ in addition to $1$. We conclude that $d/L_0 \approx 0.06$
is sufficient at the level of our accuracy. Therefore for $L_0=33$ and 
$L_0=66$ we used $d=2$ and $d=4$, respectively.
Throughout we used $L > 5 L_0$, which is clearly sufficient to neglect
deviations from the limit $L \rightarrow \infty$; See ref. \cite{mybreaking}.
We chose $\beta_0=0.15$, $0.19$, $0.21$ and $0.218$ for $L_0=8.5$, $16.5$,
$33$ and $66$, respectively. We simulated at $163$, $122$, $117$ and $41$ 
values of $\beta$ for these thicknesses, respectively. Note that in the case of
$L_0=66$ we simulated only up to $\beta_c$, since these simulation are 
rather expensive. 

For $L_0=16$ and $17$ we performed  $6.4 \times 10^8$ measurements for each 
value of $\beta$ that we simulated at. For each measurement we performed 
$16$ sweeps with the Metropolis algorithm. In total these simulations took  
the equivalent of about 8 years of CPU time on one core of a Quad-Core AMD Opteron(tm) 
Processor 2378 running at 2.4 GHz.
For $L_0=15$ and $18$  we performed $1.3 \times 10^8$ measurements
and for $L_0=14$ and $19$ only $6.4 \times 10^7$ measurements.
For $L_0=32$ we performed between $2.6 \times 10^7$ and 
$6.4 \times 10^7$ measurements and for $L_0=34$ we measured $2.6 \times 10^7$
or $3.2 \times 10^7$ times for each value of $\beta$. These 
simulations took the equivalent of about 5 years of CPU time on one core of a 
Quad-Core AMD Opteron(tm) Processor 2378 running at 2.4 GHz.
For $L_0=64$ and $68$ we performed $6.4 \times 10^6$ measurements for each 
value of $\beta$. In total these simulations took the equivalent of about 2.5 years
of CPU time on one core of a Quad-Core AMD Opteron(tm) Processor 2378
running at 2.4 GHz.

We improved the numerical results obtained in ref. \cite{mybreaking}
for the Blume-Capel model.  To this end, we simulated at additional values
of $\beta$. This way both the statistical error of our result
as well as the systematical error of the numerical integration are reduced. 
In ref. \cite{mybreaking} we simulated
the thicknesses $L_0=8$, $9$, $16$, $17$, $32$ and $33$. Here we simulated
$L_0=34$ in addition. 

\begin{figure}
\begin{center}
\includegraphics[width=14.5cm]{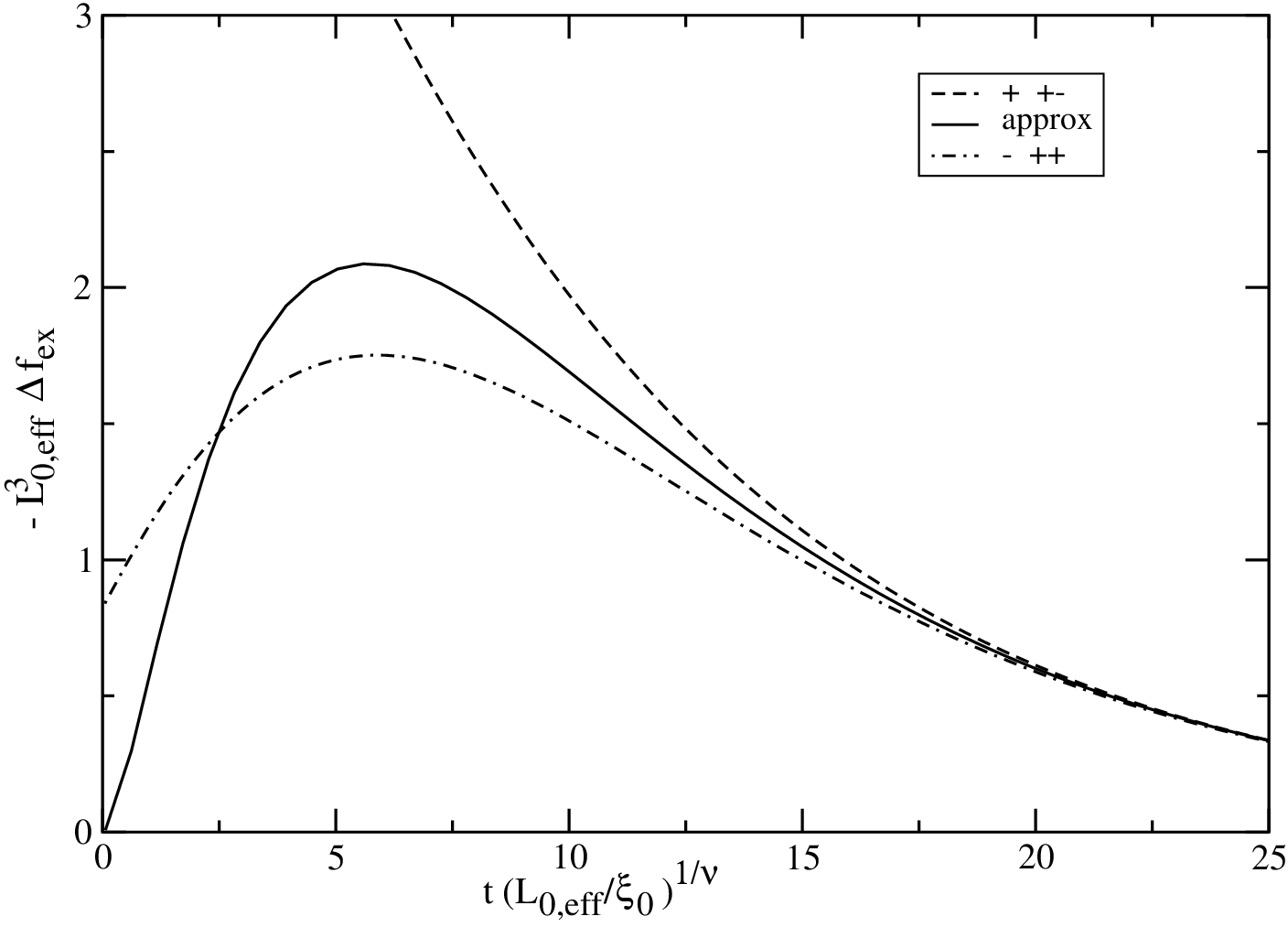}
\caption{\label{approxplot}
We plot $\theta_{+-}$, $-\theta_{++}$ and the 
approximation~(\ref{startingpoint}). The data are taken for the Blume-Capel
model at $D=0.655$ and the finite difference is computed from $L_0=32$ and 
$L_0=34$.
}
\end{center}
\end{figure}

In figure \ref{approxplot} we  plot $\theta_{+-}$, $-\theta_{++}$
and the approximation~(\ref{startingpoint}) computed by using the data 
obtained for the Blume-Capel model at $D=0.655$ for $L_0=33$ and $d=2$.
As discussed at the end of section \ref{Dffff}, 
we used the value $L_s=1.91$ to compute the effective thickness
$L_{0,eff}=L_0+L_s$. The deviation
of $\theta_{+-}$ and $-\theta_{++}$ from the approximation~(\ref{startingpoint})
is smaller than $5 \%$ for $x \gtrapprox 16$   and smaller than $1 \%$ 
for $x \gtrapprox 22.5$. The average $(\theta_{+-} -\theta_{++})/2$ 
deviates from the approximation~(\ref{startingpoint}) by less than $5 \%$
for $x \gtrapprox 8.6$ and by less than $1 \%$  for $x \gtrapprox 12.7$. 

\begin{figure}
\begin{center}
\includegraphics[width=14.5cm]{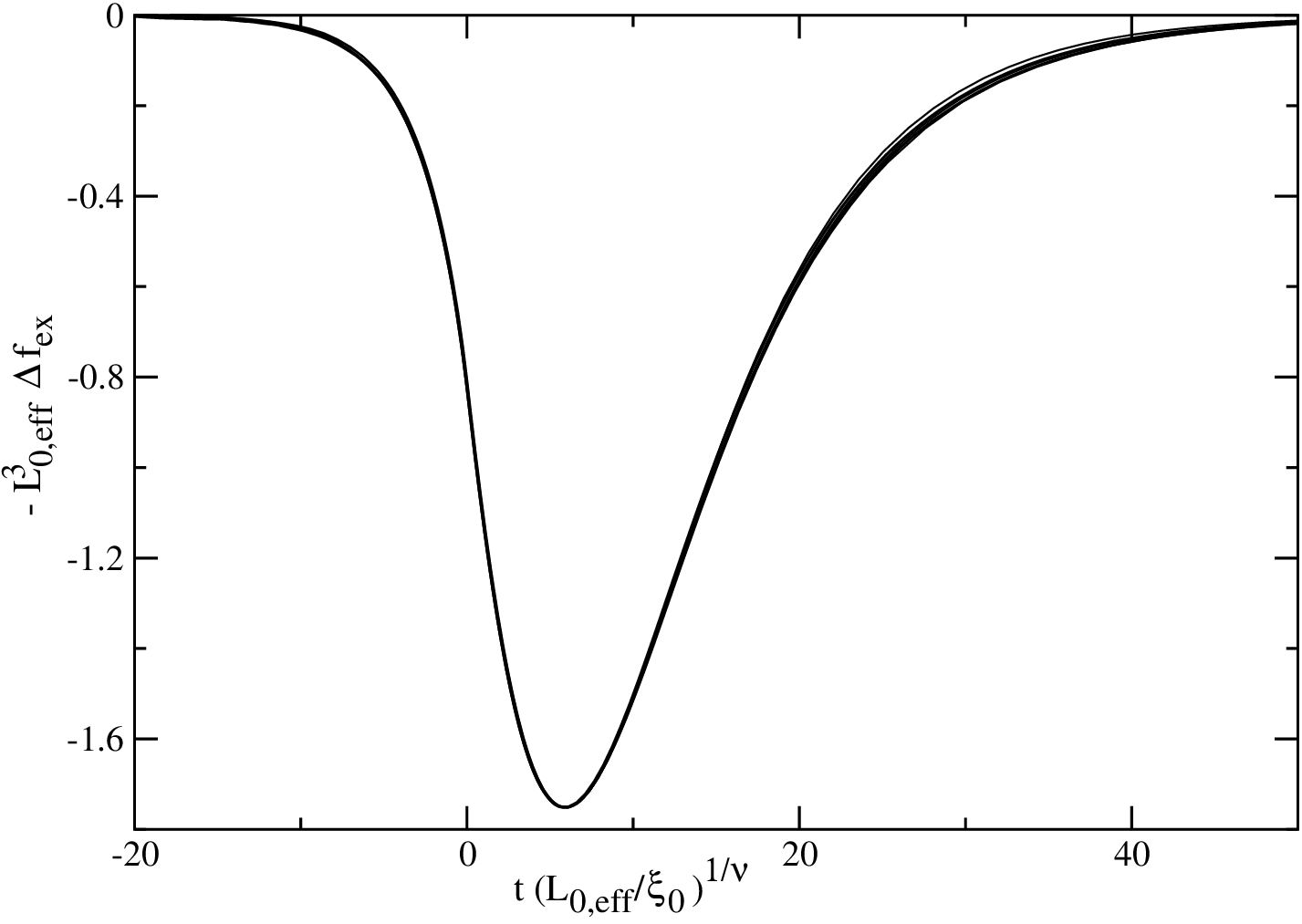}
\caption{\label{thetapp}
We plot $-\Delta f_{ex} L_{0,eff}^3$ as a function of 
$t [L_{0,eff}/\xi_{0}]^{1/\nu}$ for $++$ boundary conditions.
The thick lines give the result obtained for the Blume-Capel model at $D=0.655$
and the two thicknesses $L_0=16.5$ and $L_0=33$. In the case of the Blume-Capel model we used $L_{0,eff}=L_0 + 1.91$ as effective thickness of the 
film. Our results for the Ising model are given by thin lines.
In the case of the Ising model we used the effective thicknesses
$L_{0,eff}=19.712$, $L_{0,eff}=36.509$ and 
$L_{0,eff}=69.936$, for $L_0=16.5$, $L_0=33$ and $L_0=66$, respectively.
These effective thicknesses are chosen such that at the minima the curves
fall on top of the one for the Blume-Capel model and $L_0=33$. 
At the resolution of the plot, all 5 curves fall on top of each other almost
everywhere. Only for $20 \lessapprox  x  \lessapprox 40$   the curve for 
the Ising model and $L_0=16.5$ can be distinguished from the other four.
}
\end{center}
\end{figure}

Next we extracted the value and the location of the minimum of 
$- \Delta f_{ex,++}$.  In the case of the Blume-Capel model we get 
$\beta_{min} =0.382185(15)$ and $- \Delta f_{ex,++,min}=-0.0002808(6)$ 
for $L_0=16.5$  and 
$\beta_{min} =0.385716(6)$ and $- \Delta f_{ex,++,min}=-0.00004117(5)$ for $L_0=33$.
This corresponds to $t_{min} [L_{0,eff}/\xi_0]^{1/\nu}=5.88(5)$ and 
$- L_{0,eff}^3 \Delta f_{ex,++,min}= -1.752(18)$  for $L_0=16.5$ and 
$t_{min} [L_{0,eff}/\xi_0]^{1/\nu}=5.88(4)$ and $- L_{0,eff}^3 \Delta f_{ex,++,min}= 
-1.752(10)$ for $L_0=33$. The quoted error-bars include the error of
$\beta_{min}$, $- \Delta f_{ex,++,min}$ and errors induced by the uncertainties 
of $L_s$, $\xi_0$, $\nu$ and $\beta_c$. The values obtained from $L_0=16.5$
and  $L_0=33$ agree nicely. Our results are also consistent with those of
ref. \cite{mybreaking}: $x_{min} = 5.82(10)$ and
$\theta_{++,min}=-1.76(3)$. Our results obtained for the Ising model are
summarized in table \ref{Isingppmin}. Here we computed $L_{0,eff}$ 
by requiring $- L_{0,eff}^3 \Delta f_{ex,++,min} = -1.75169...$ which is our 
estimate obtained for the Blume-Capel model and $L_0=33$. 
We see that the values of $L_{0,eff}$ are similar to those obtained in 
section \ref{fssY} from the analysis of the free energy differences
at the critical point. In the last
column we give $t_{min} [L_{0,eff}/\xi_0]^{1/\nu}$ using these values of
$L_{0,eff}$. We see that these estimates of $x_{min}$ are essentially
consistent with that obtained above from the analysis of the Blume-Capel
model. 

For $L_0=16.5$ we checked the effect of the discretization error
on the position and the value of the minimum. The error behaves as
$\epsilon=a d^2 + \mbox{O}(d^4)$. The results obtained for $d=1$, 
$3$ and $5$
are consistent with a quadratic behaviour. For $d=1$, the relative error 
is about one permille for both $- \Delta f_{ex,++,min}$ and $t_{min}$.

In figure \ref{thetapp} we plot our numerical results for the 
scaling function $\theta_{++}$ which are given by 
$- L_{0,eff}^3 \Delta f_{ex,++}$ as a function of $t [L_{0,eff}/\xi_0]^{1/\nu}$
where $\nu=0.63002$ is set.
In the case of the Blume-Capel model we  use $L_{0,eff}=L_0 + 1.91$ 
as effective thickness of the film. We give our results for $L_0=16.5$
and $33$. 
For the Ising model we  take the effective thicknesses
given in the sixth column of table  \ref{Isingppmin}. We plot 
our results for $L_0=16.5$, $d=1$, $L_0=33$ and  $L_0=66$.
The error bars are too small to be visible in the plot.
At the resolution of the plot, all 5 curves fall on top of each other almost
everywhere. Only for $20 \lessapprox  x  \lessapprox 40$  the curve for
the Ising model and $L_0=16.5$ can be distinguished from the other four.

\begin{table}
\caption{\sl \label{Isingppmin}  Results for the minimum of $\theta_{++}$ 
obtained for the Ising model. 
}
\begin{center}
\begin{tabular}{ccclcc}
\hline
$L_0-d/2$& $L_0+d/2$ & $\beta_{min}$ &\mc{1}{c}{$-\Delta f_{ex,++,min}$} & $L_{0,eff}$ & 
$t_{min} [L_{0,eff}/\xi_0]^{1/\nu}$ \\
\hline
 8 & 9   & 0.2123025(16)  &  --1.1605(1) $\times  10^{-3}$ & 11.471 & 5.96(1) \\
14& 19   & 0.2176215(5)   &  --2.347(1)  $\times  10^{-4}$ &        &\\
15 &18   & 0.2176744(19)  &  --2.306(1)  $\times  10^{-4}$ &        &\\
16 &17   & 0.2176975(30)  &  --2.2869(15) $\times 10^{-4}$ & 19.712 & 5.96(1) \\
32 &34   & 0.2201704(30)  &  --3.5996(26) $\times 10^{-5}$ & 36.509 & 5.94(2) \\
64 &68   & 0.2211284(25)  &  --5.121(18)  $\times 10^{-6}$ & 69.936 & 5.91(3) \\
\hline
\end{tabular}
\end{center}
\end{table}

\begin{figure}
\begin{center}
\includegraphics[width=14.5cm]{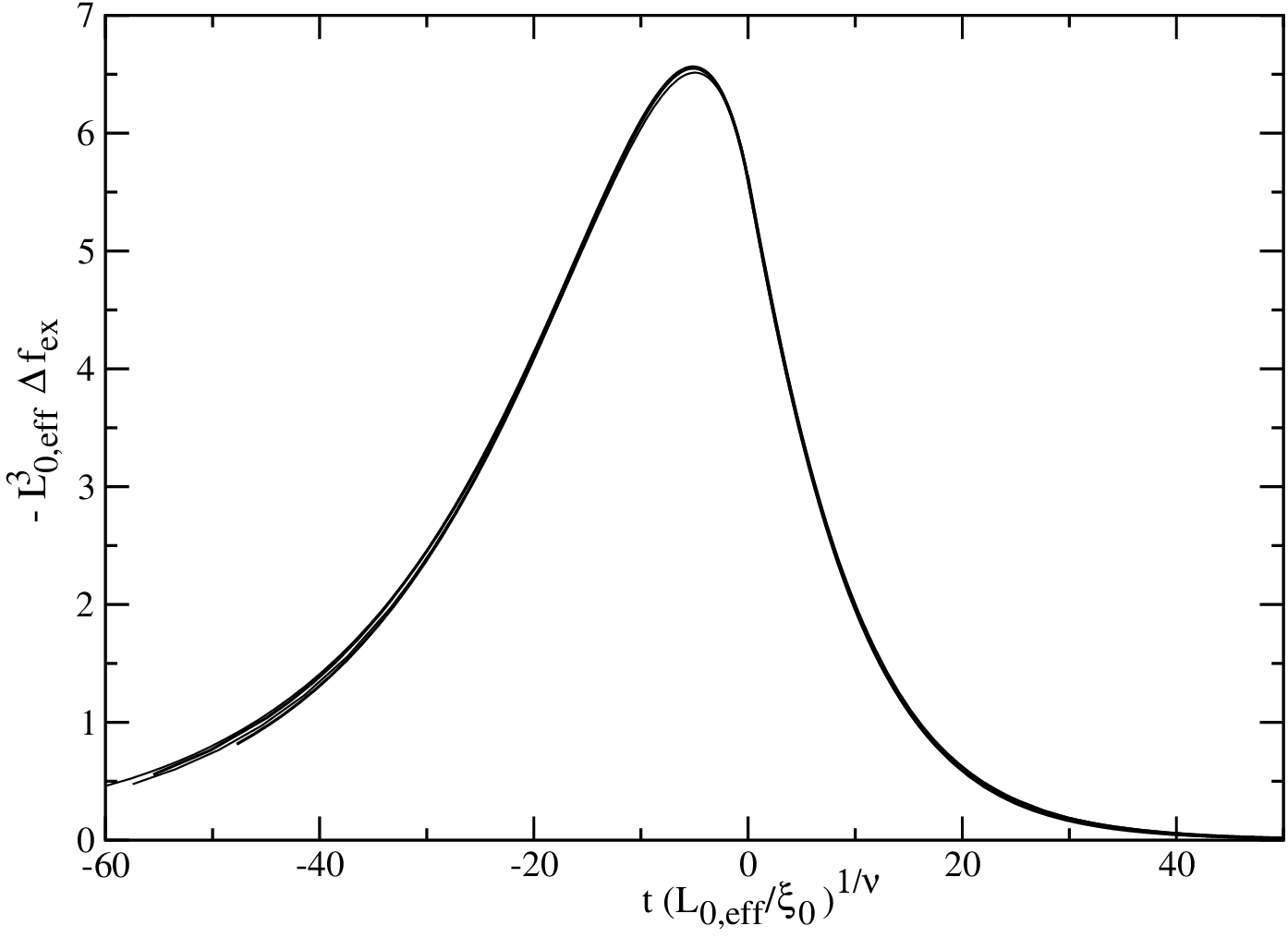}
\caption{\label{thetapm}
We plot $- L_{0,eff}^3 \Delta f_{ex}$ as a function of $t [L_{0,eff}/\xi_{0}]^{1/\nu}$
for $+-$ boundary conditions.
The thick lines give the result obtained for the Blume-Capel model at $D=0.655$
and the two thicknesses $L_0=16.5$ and $L_0=33$. In the case of the Blume-Capel model we used $L_{0,eff}=L_0 + 1.91$ as effective thickness of the
film.
 Our results for the Ising model are given by thin lines.
In the case of the Ising model we used the effective thicknesses
$L_{0,eff}=19.712$, $L_{0,eff}=36.509$ and
$L_{0,eff}=69.936$, for $L_0=16.5$, $L_0=33$ and $L_0=66$, respectively.
These values are taken from the analysis of $++$ boundary conditions
above. At the resolution of the plot, all 5 curves fall on top of each other 
almost
everywhere. Near the maximum the curve for the Ising model and $L_0=16.5$ stays
slightly below the other ones. For $x \lessapprox -30$ the curves slightly fork.
Note that in this range the difference between the Blume-Capel results for 
$L_0=16.5$ and $L_0=33$ is of a similar size as the one between the Ising results for
$L_0=16.5$ and $L_0=33$ and between Blume-Capel and Ising.
}
\end{center}
\end{figure}

Next we discuss our numerical results for the scaling function $\theta_{+-}$. 
In figure \ref{thetapm} we plot
$- L_{0,eff}^3 \Delta f_{ex,+-}$ as a function of $t [L_{0,eff}/\xi_{0}]^{1/\nu}$ for the 
Blume-Capel model at the thicknesses $L_0=16.5$ and $33$ and the Ising model
at $L_0=16.5$, $33$ and $66$.  In the case of the Blume-Capel model we use 
$L_{0,eff}=L_0+L_s$ with $L_s=1.91$. For the Ising model we take the same 
values for $L_{0,eff}$ as above for $++$ boundary conditions.

We find that the different curves fall quite nicely on top of each other. 
In the neighbourhood of the maximum the curve for the Ising model at $L_0=16.5$
lies slightly below the other ones and for $x \lessapprox -30$ the 
curves slightly fork.
The discrepancies discussed for $++$ boundary conditions in the range
$20 \lessapprox  x  \lessapprox 40$ are also present for $+-$ boundary 
conditions.
They can not be seen in figure \ref{thetapm} since the range of values for 
$+-$ boundary conditions is larger than that for $++$ boundary conditions.

In table \ref{Isingpmmax} we summarize results for the maximum of 
$\theta_{+-}$. Using $L_s=1.91$ in the case of the 
Blume-Capel model we get nicely consistent results for $x_{max}$ and 
$\theta_{+-,max}$ from the two thicknesses $L_0=16.5$ and  $L_0=33$. 
These results improve those of ref. \cite{mybreaking}: $x_{+-,max}=-5.17(7)$
and $\theta_{+-,max}=6.56(10)$.  In the case of the Ising model we 
use the values of $L_{0,eff}$ obtained above for films with 
$++$ boundary conditions. The resulting estimates for $x_{max}$ and 
$\theta_{+-,max}$ are close to those obtained from the Blume-Capel model.
In particular the results obtained for $L_0=33$ are closer to the 
Blume-Capel ones than those obtained for $L_0=16.5$.

We conclude that our numerical results for the scaling functions of the 
thermodynamic Casimir force for $++$ and $+-$ boundary conditions are
fully consistent with the universality hypothesis. Furthermore
our ansatz~(\ref{leff2}) provides a good approximation of the universal
correction function. 

\begin{table}
\caption{\sl \label{Isingpmmax}  Results for the maximum of $\theta_{+-}$
obtained for Blume Capel (BC) model and the Ising (I) model. In the second and 
third column we give the thicknesses that have been considered. In the 
fourth column we give the value of $-\Delta f_{ex,+-}$ at the maximum and in the 
fifth column we give the location of the maximum. In the sixth and seventh
column we give estimates of $\theta_{+-,max}$ and $x_{max}$ derived from these
results. 
}
\begin{center}
\begin{tabular}{cccllll}
\hline
Model & $L_0-d/2$& $L_0+d/2$ & \mc{1}{c}{$\beta_{max}$} 
&\mc{1}{c}{$-\Delta f_{ex,+-,max}$} & 
\mc{1}{c}{$-L_{0,eff}^3 \Delta f_{ex,+-,max}$} & 
\mc{1}{c}{$t_{max} [L_{0,eff}/\xi_0]^{1/\nu}$} \\
\hline
 BC & 16 & 17&  0.39257(3) & 0.0010501(7)    &  6.552(5)[54] &--5.15(3)[3] \\
 BC & 32 & 34&  0.389474(5) & 0.00015426(5)  &  6.563(2)[28] &--5.139(15)[15]\\
\hline
 I & 16 & 17  & 0.224948(4) & 0.00085044(30)  & 6.514(2) &--4.959(6)  \\
 I & 32 & 34  & 0.2229119(3)& 0.000134650(35) & 6.552(2) &--5.035(12) \\
\hline
\end{tabular}
\end{center}
\end{table}

\section{Summary and Conclusions}
We studied the spin-1/2 Ising model and the improved Blume-Capel
model on the simple cubic lattice with film geometry. In particular 
we  considered strongly symmetry breaking $++$ and $+-$ 
boundary conditions. We focused on the thermodynamic Casimir force.
At the critical point we  studied the behaviour of the 
free energy per area, the energy per area, the magnetisation profile
and the second moment correlation length of the film.
The main subject of the present work are corrections to scaling.
Previously it has been demonstrated at the example of improved models
that corrections $\propto L_0^{-1}$ that are due to the boundaries 
can be expressed by an effective thickness $L_{0,eff}=L_0 + L_s$, 
where $L_s$ is the same for all quantities. Note that $L_s$ depends on the 
model and in particular on the details of the boundary conditions.
Here we probed the hypothesis that the leading bulk corrections
can be expressed in an analogous way:
\begin{equation}
\label{leffnochmal}
 L_{0,eff} = L_0 + L_s + c (L_0 + L_s)^{1-\omega} \;\;.
\end{equation}
Fitting various quantities at the critical point of the Ising model
we find similar, but likely not identical values of the amplitude
$c$. Also the study of the thermodynamic Casimir force for large
values of the scaling variable $x$ shows that eq.~(\ref{leffnochmal})
can not be exact. Nethertheless it turns out to be a surprisingly 
good approximation in the range of $x$ that is of experimental 
interest. In section \ref{thetafull} we investigate the thermodynamic 
Casimir force for $++$ and $+-$ boundary conditions. 
We find for $- L_{0,eff}^3 \Delta f_{ex}$ plotted
as a function of $t [L_{0,eff}/\xi_0]^{1/\nu}$ a good collapse of the 
data for both the spin-1/2 Ising model and the Blume-Capel model.
In the case of the Blume-Capel model we used $L_{0,eff}=L_0+L_s$
with $L_s=1.91(5)$. We demonstrated that in the case of the 
spin-1/2 Ising model approximately the same $L_{0,eff}$ can be used 
for $++$ and $+-$ boundary conditions. The values of $L_{0,eff}$ that
we obtained in section \ref{thetafull} for $L_0=16.5$, $33$ and 
$66$ are similar to those obtained from the analysis of $D_{f,+-,++}$ 
in section \ref{Dffff}. The estimates of $L_s$ and $c$ obtained from 
this analysis are highly anti-correlated. From the analysis of 
$D_{f,+-,++}$ we get $L_s=0.9$ and $c=1.5$ as central estimates.
The range of possible values is given by $L_s=1.1$, $c=1.4$
one side and $L_s=0.8$, $c=1.6$ at the other.  Note that the value
of $L_s$ depends on the definition of the thickness. In particular, 
when comparing with refs. \cite{VaGaMaDi07,VaGaMaDi08,VaMaDi11} (VGMD) one 
should take into account that $L_{0,VGMD}=L_{0,ours}+2$ and hence 
$L_{s,VGMD}=L_{s,ours}-2$. Since the correction function $q(x)$ is 
universal, also for experimental data or data obtained from the numerical
study of other models an effective thickness~(\ref{leffnochmal}) should
parametrize leading corrections quite well. Note again that $L_s$ should
depend on the microscopic details of the system. 
In the case of the amplitude
$c$ universal ratios can be constructed. For example
\begin{equation}
 \frac{c}{a_{\xi,+} \xi_0^{\omega} } = -8(2) \;\; 
\end{equation}
where we used the numerical values of $a_{\xi,+}$ and $\xi_0$ obtained
in the Appendix.  In the introduction we argued that eq.~(\ref{leff2})
provides a good approximation for the corrections to scaling function 
since fluctuations
are strongly suppressed near the boundaries of the film. Therefore 
eq.~(\ref{leff2}) should not work for periodic and anti-periodic 
boundary conditions. Furthermore the amplitude of leading corrections should
be smaller in these cases, which is indeed confirmed by the numerical
results of \cite{HuGrSc11} for periodic boundary conditions.

Furthermore we improved the numerical accuracy of the estimates
of the universal scaling functions $\theta_{++}$ and $\theta_{+-}$:

Writing the partition function in terms of eigenvalues and 
eigenstates of the transfer matrix and boundary states one finds for 
large values of $x$
\begin{equation}
 \theta_{++}(x) = - \theta_{+-}(x) = - C^2  x^{3 \nu} \exp(-x^{\nu}) \;\;.
\end{equation}
Here we demonstrated how $C^2$ can be accurately computed by analysing 
the magnetisation profile of films and the bulk correlation function.
We find 
\begin{equation}
C^2 =1.552(2) \;\;.
\end{equation}
This result can be compared with $C^2= 1.5(1)$ obtained in 
ref. \cite{mybreaking}.

At the critical point we find by studying the difference of free energies
per area
\begin{equation}
\Delta_{+-}- \Delta_{++} =  [\theta_{+-}(0)-\theta_{++}(0)]/2 =3.204(5)
\end{equation}
where we average the results obtained from the analysis of the 
spin-1/2 Ising and the improved Blume-Capel model. For the slope
of the scaling function at the critical point we find
\begin{equation}
 \theta'_{+-}(0) =-0.482(2) \;\;,\;\; \theta'_{++}(0) =-0.318(2) \;\;.
\end{equation}

The minimum of $\theta_{++}$ is located at $x_{min}=5.88(4)$
and takes the value $\theta_{++,min}=-1.752(10)$. For the maximum of
$\theta_{+-}$ we get $x_{max}=-5.14(3)$ and $\theta_{+-,max}=6.56(3)$.
The reduction of the  error compared with ref. \cite{mybreaking} 
is mainly due to the fact that here we assume $L_s=1.91(5)$ instead of
$L_s=1.9(1)$ as in ref. \cite{mybreaking}.

\section{Acknowledgements}
This work was supported by the DFG under the grant No HA 3150/2-2.

\appendix
\section{Numerical results for the spin-1/2 Ising bulk system}
\subsection{The critical point}
\label{appendixbc}
We extended the study of ref. \cite{mycritical} by simulating the 
Ising model on the simple cubic lattice on a system of the size $L^3$ with 
$L=400$ and periodic boundary conditions in all three directions 
at $\beta=0.2216546$. As in ref. \cite{mycritical} we simulated 
the model by using a hybrid of the local Metropolis algorithm, the single
cluster algorithm \cite{Wolff} and the wall cluster algorithm \cite{wall}. 
For details see section IV of ref. \cite{mycritical}. We performed 
$2.3 \times 10^7$ measurements. In total this simulation took  
the equivalent of about 4 years of CPU time on 
a single core of a  Quad-Core AMD Opteron(tm) Processor 2378 running at 2.4 GHz.
In the first step of the analysis we determined 
$\beta_c$  by analysing the behaviour of the renormalization group invariant
quantities $Z_a/Z_p$, $\xi_{2nd}/L$, $U_4$ and $U_6$. For the definition of 
these quantities see section II of ref. \cite{mycritical}. We  
fitted our data for the Ising model with the ansatz
\begin{equation}
 R(\beta_c,L) = R^* + a L^{-\omega} + b L^{-2} 
\end{equation}
where $R$ denotes one of the renormalization group invariant quantities. 
Performing these fits, we  used the results for $R^*$ given in table V
of ref. \cite{mycritical} as input. Furthermore, we  fixed $\omega=0.832$. 
We get acceptable $\chi^2/$d.o.f. for fits with $L_{min} \ge 16$. The
statistical error of $\beta_c$ increases only slowly with increasing $L_{min}$.
Based on fits with $L_{min} \ge 24$ 
for $Z_a/Z_p$ and $\xi_{2nd}/L$ we arrive at $\beta_c=0.22165462(2)$. Instead, 
analysing $U_4$ and $U_6$ we arrive at $\beta_c=0.22165463(2)$. 
In ref. \cite{KaRiMe01}
the authors computed the Binder cumulant $U_4$ on lattices of a linear 
size up to $L=1536$.  Fitting their data, taking
the value $U_4^*=1.6036(1)$  \cite{mycritical} as input, we arrive
at $\beta_c= 0.221654615(10)$.  In this work we shall use  
\begin{equation}
\label{betacI}
\beta_c =0.22165462(2) \;\;.
\end{equation} 
This estimate can be compared e.g. with the previous estimates 
$\beta_c = 0.22165463(8)$ obtained in ref. \cite{mycritical} using 
a linear lattice size up to $L=96$  and $\beta_c= 0.22165455(3)$ given in 
table X of \cite{DeBl03X}.

At the critical point the energy density behaves as 
\begin{equation}
\label{energybc}
 E_{bulk}(L) = E_{ns}  + a L^{3-1/\nu} \times (1 + b L^{-\omega} + ...) \;\;.
\end{equation}
Performing various fits based on eq.~(\ref{energybc}), using the data
of ref.~(\cite{mycritical}) and our result for $L=400$, we arrive at 
\begin{equation}
\label{Enonsingular}
 E_{ns} = 0.9906065(15) + 85 \times (\beta_c - 0.22165462) 
\end{equation}
The specific heat behaves as 
\begin{equation}
\label{heatbc}
 C_{bulk}(L) = C_{ns}  + a L^{3-2/\nu} \times (1 + b L^{-\omega} + ...)
\end{equation}
performing various fits based on eq.~(\ref{energybc}), using the data
of ref.~(\cite{mycritical}) and our result for $L=400$, we arrive at
\begin{equation}
\label{Cnonsingular}
 C_{ns} = -29.1(3) - 7700000 \times (\beta_c - 0.22165462)
                  - 3300 \times (\nu- 0.63002) \;\;.
\end{equation}

\subsection{Amplitudes and amplitude ratios}
\label{appendixamp}
We simulated the three-dimensional Ising model for a large number of 
$\beta$-values in the high and the low temperature phase on 
$L^3$ lattices with periodic boundary conditions in all three directions.
We have chosen the linear lattice size such that $L > 10 \xi_{2nd}(\beta)$ 
in order to keep deviations from the thermodynamic limit sufficiently small
to be ignored in the analysis of the data.
For the precise definition of the observables see section II of  
\cite{myamplitude}.
In the high temperature phase we simulated at 68 values of $\beta$ in the 
range $0.125 \le \beta \le 0.2213$. To give the reader an impression of the 
quality of the data, we give the results for the 5 largest values of $\beta$ 
in table \ref{reshigh}. Analogous results for the low temperature phase are 
given in table \ref{reslow}.

\begin{table}
\caption{\sl \label{reshigh} The second moment correlation length $\xi_{2nd}$, 
the magnetic susceptibility $\chi$ and the energy density $E_{bulk}$ 
for the five largest values of the inverse temperature $\beta$
that we simulated in the high temperature phase of the Ising model.
We  simulated $L^3$ systems with periodic boundary conditions in all 
three directions.
}
\begin{center}
\begin{tabular}{cclll}
\hline
$\beta$  & $L$ &\mc{1}{c}{$\xi_{2nd}$}  & \mc{1}{c}{$\chi$} 
&\mc{1}{c}{$E_{bulk}$} \\
\hline
  0.2206 & 200 &  14.57699(31) &\phantom{0}831.162(32) &  0.96369936(90) \\  
  0.2207 & 200 &  15.5321(10)  &\phantom{0}940.79(11)  &  0.9656874(29)  \\
  0.2208 & 200 &  16.6644(11)  &          1079.27(14)  &  0.9677195(31)  \\
  0.2210 & 300 &  19.73548(63) &          1501.960(86) &  0.97198710(87) \\
  0.2213 & 400 &  29.1058(11)  &          3212.44(23)  &  0.97909806(69) \\
\hline
\end{tabular}
\end{center}
\end{table}

\begin{table}
\caption{\sl \label{reslow} 
The second moment correlation length $\xi_{2nd}$,
the magnetic susceptibility $\chi$, the magnetisation $m$ and the energy 
density $E_{bulk}$ for the five smallest values of the inverse 
temperature $\beta$
that we simulated in the low temperature phase of the Ising model.
We simulated $L^3$ systems with periodic boundary conditions in all three
directions.
}
\begin{center}
\begin{tabular}{cccccc}
\hline
$\beta$  & $L$ &  $\xi_{2nd}$  &  $\chi$     &  $m$ &   $E_{bulk}$ \\
\hline
  0.2219 & 300 &  18.930(40)&          1058.49(66)& 0.1815607(39) & 1.0126483(10) \\
  0.2220 & 200 &  15.294(24)&\phantom{0}690.78(38)& 0.2027298(54) & 1.0200656(17) \\
  0.2221 & 200 &  12.976(28)&\phantom{0}501.95(30)& 0.2200006(48) & 1.0271260(16) \\
  0.2222 & 170 &  11.418(17)&\phantom{0}389.43(17)& 0.2347800(43) & 1.0339257(16) \\
  0.2223 & 170 &  10.278(13)&\phantom{0}315.26(12)& 0.2477779(38) & 1.0405068(16) \\
\hline
\end{tabular}
\end{center}
\end{table}

First we fitted our data for the second moment correlation length in the 
high temperature phase using the ansaetze
\begin{equation}
\label{xifit1}
 \xi_{2nd} = \xi_{2nd,0,+} t^{-\nu} \times (1 + a_{\xi,+} t^{\theta} )
\end{equation}
\begin{equation}
\label{xifit2}
 \xi_{2nd} = \xi_{2nd,0,+} t^{-\nu} \times (1 + a_{\xi,+} t^{\theta} + b t )
\end{equation}
and
\begin{equation}
\label{xifit3}
 \xi_{2nd} = \xi_{2nd,0,+} t^{-\nu} \times (1 + a_{\xi,+} t^{\theta} + b t + c t^{2 \nu} )
\end{equation}
where $t = \beta_c -\beta$. We fixed  $\beta_c = 0.22165462$, $\nu=0.63002$
and $\omega=0.832$.  
Based on a large number of fits using these ansaetze we conclude
\begin{equation}
\label{amplitudexi}
 \xi_{2nd,0,+}= 0.1962(1) +  540 \times (\beta_c-0.22165462) 
                      -1.8 \times  (\nu-0.63002)
                      -0.002 \times (\omega-0.832)
\end{equation} 
and 
\begin{equation}
\label{correctionaxi}
 a_{\xi,+}= -0.32(3) - 120000 \times (\beta_c-0.22165462)
           + 130  \times  (\nu-0.63002)
            -1.1  \times (\omega-0.832) \;\;.
\end{equation}
Our result is in nice agreement with that of ref. \cite{BuPe11} obtained 
by analysing the high temperature series of $\xi_{2nd}$.  In table VII of 
\cite{BuPe11} the authors quote $\xi_{0,+} = 0.5070(5)$ for the definition
$\tilde t = (\beta_c-\beta)/\beta_c$ of the reduced temperature.  Converting 
to our convention one gets 
$\xi_{0,+} = 0.5070(5) \times 0.22165462^{0.63002}=0.1962(2)$.

In a similar way we analysed the second moment correlation length 
in the low temperature phase and the magnetic susceptibility in both 
phases. Let us summarize the final results:
\begin{equation}
\label{amplitudexilow}
 \xi_{2nd,0,-} = 0.1015(2)    -200  \times (\beta_c-0.22165462)
                          - 0.9 \times (\nu-0.63002) 
                          -0.001 \times (\omega-0.832)
\end{equation}
and
\begin{equation}
 a_{\xi,-} = -0.55(15) + 70000  \times (\beta_c-0.22165462)
                       +100 \times (\nu-0.63002)
               -2.2 \times (\omega-0.832)
\end{equation}
Using the results~(\ref{amplitudexi}) and (\ref{amplitudexilow}) we get
for the universal ratio $\xi_{2nd,0,+}/\xi_{2nd,0,-} =  1.933(5)$, 
which is fully 
consistent with $\xi_{2nd,0,+}/\xi_{2nd,0,-} = 1.939(5)$ obtained in ref. 
\cite{myamplitude} by analysing Monte Carlo data obtained for the 
Blume-Capel model at $D=0.655$. 

Analysing the data for the magnetic susceptibility in the high 
temperature phase we arrive at
\begin{equation}
 C_+ =  0.1739(1)  + 800 \times (\beta_c-0.22165462)
                   -1.6  \times (\gamma-1.2372)
                   -0.0013  \times (\omega-0.832)
\end{equation}
and
\begin{equation}
a_{\chi,+} = -0.33(5)  -150000 \times (\beta_c-0.22165462)
                       + 100 \times (\gamma-1.2372)
                      -1.3 \times (\omega-0.832) \;.
\end{equation}

The corresponding result for the low temperature phase are
\begin{equation}
 C_- = 0.03695(2) - 200 \times (\beta_c-0.22165462)
                  -0.35 \times (\gamma-1.2372)
                  -0.001 \times (\omega-0.832)
\end{equation}
and
\begin{equation}
a_{\chi,-} = -1.6(2) + 20000 \times (\beta_c-0.22165462)
                     + 120  \times (\gamma-1.2372)
                     -  7 \times (\omega-0.832) \;.
\end{equation}
The ratio $C_+/C_-= 4.706(8)$ is consistent with $C_+/C_-=4.713(7)$ 
obtained in ref. \cite{myamplitude} by analysing Monte Carlo data obtained 
for the Blume-Capel model at $D=0.655$. Note that our estimates are slightly
smaller than $C_+/C_-=4.78(3)$ obtained from series expansions \cite{BuPe11}.

\subsection{The energy density}
\label{appendixene}
In order to compute the thermodynamic Casimir force, we need the energy
density of the bulk system for a large number of $\beta$ values. To
this end, the authors of ref. \cite{HuGrSc11} used the results of
of ref. \cite{Feng} in combination with a naive evaluation of 
the high \cite{ArFu03} and low \cite{BhCHLW94} temperature series.
Here, instead, we 
combined the analysis of the high \cite{ArFu03} and low \cite{Vohwinkel} 
temperature series with the results of our Monte Carlo simulations 
discussed above. The analysis of the high temperature series is simpler
and the results are more accurate than that of the low temperature one.
This is due to the fact that the high temperature series converges up to 
the critical point, while this is not the case for the low temperature
series. 

In the neighbourhood of the 
critical point the energy density behaves as
\begin{equation}
\label{Ebehave}
 E_{bulk} = E_{ns} - C_{ns} t + ... 
   + a_{\pm} |t|^{1-\alpha} (1 + b_{\pm} |t|^{\theta} + ...)
   + ...
\end{equation}
We analysed both series using differential approximants. In particular, 
we used the second order differential equation given in eq.~(6.16) of 
ref. \cite{Guttmann}: 
\begin{equation}
\label{diffequation}
 u^2 Q_2(u) g''(u) + u Q_1(u) g'(u) + Q_0(u) g(u) =R(u)
\end{equation}
where $Q_2(u)$, $Q_1(u)$, $Q_0(u)$ and $R(u)$ are polynomials in the expansion 
variable $u$ of the order $J$, $K$, $L$ and $M$, respectively. These polynomials
are fixed by the requirement that the function $g(u)$ has the correct expansion
in $u$ up to the highest known order. The differential eq.~(\ref{diffequation})
is used, since it is known that its solution  behaves as
\begin{equation}
 g(u) = g_{ns}(u) + a_1(u) (u_c-u)^{-x_1}  + a_2(u) (u_c-u)^{-x_2}
\end{equation}
where $g_{ns}(u)$, $a_1(u)$ and $a_2(u)$ are analytic functions.

Usually one sets $Q_2(0)=1$.  Therefore
$J+K+L+M=N-2$, where $N$ is the order of the last known coefficient of the 
series. We biased the analysis by using our
estimate~(\ref{betacI}) of the inverse critical temperature and our 
estimates of $\nu$ and $\omega$ \cite{mycritical}. This way additional 
coefficients of the polynomials are  fixed and one gets $J+K+L+M=N+3$. 
For a detailed discussion we refer the reader to section 6 of 
ref. \cite{Guttmann}.
We solved the differential equation~(\ref{diffequation}) numerically by 
using the  Runge-Kutta method.

In the high temperature phase Arisue and Fujiwara  \cite{ArFu03}
computed the free energy density of the bulk system as a series 
in $v=\tanh(\beta)$ up to $O(v^{46})$. Note that the coefficients of  
odd orders vanish and hence the free energy density can be expressed as 
a series in $u=v^2=\tanh^2(\beta)$. Since we are aiming at the energy density, 
we actually analysed 
\begin{equation}
 \tilde E  = - \frac{\partial f}{\partial u}  \;\;.
\end{equation}
The energy density is then given by
\begin{equation}
 E_{bulk} = - \frac{\partial f}{\partial \beta} = 
     - \frac{\partial f}{\partial u} \frac{\partial u}{\partial \beta}
   = 2 \tanh(\beta)  [1 - \tanh^2(\beta)] \; \tilde E
\end{equation}

The free energy density is given by
\begin{equation}
 -f(\beta) =\ln 2 + 3 \ln(\cosh(\beta)) + \sum_{i=0}^{46} a_i v^i + O(v^{48})
\end{equation}
where the coefficients $a_i$ are given in table I of the preprint version
of ref. \cite{ArFu03}.

We computed 
$\chi^2 = \sum_i [(E_{series}(\beta_i) - E_{MC}(\beta_i))/e(\beta_i)]^2$,
where $E_{series}(\beta_i)$ and $E_{MC}(\beta_i)$ are the estimates obtained 
from the analysis of the series and from the Monte Carlo simulations,
respectively, and $e(\beta_i)$ is the statistical error of the Monte Carlo 
result at the inverse temperature $\beta_i$.
We find that a large fraction of the possible choices of $J$, $K$, $L$ and 
$M$ result in a $\chi^2$/d.o.f.$\approx 1.03$.  About $91 \%$ of the possible 
choices have  $\chi^2/$d.o.f.$< 1.073$  and  about $92.5 \%$ 
have $\chi^2/$d.o.f.$ < 1.305$. 

We computed numerically $E_{ns}$, $C_{ns}$, $a_+$ and $a_+ b_+$ as 
defined by eq.~(\ref{Ebehave}).  Averaging over all choices of $J$, $K$, $L$ and
$M$ with $\chi^2/$d.o.f.$< 1.073$ we get
\begin{eqnarray}
 E_{ns} = 0.9906058(8) 
&+& 32 \times (\beta_c- 0.22165462) \nonumber \\
                       &-&0.0069 \times (\nu-0.63002) \nonumber \\
                       &+&0.0000072  \times (\omega-0.832) \;,
\end{eqnarray}
\begin{eqnarray}
 C_{ns} =-29.07(3) &-& 234000 \times (\beta_c- 0.22165462) \nonumber \\
                   &-&1960 \times (\nu-0.63002) \nonumber \\
                   &-&0.86  \times (\omega-0.832)  \;,
\end{eqnarray}
\begin{eqnarray}
a_+ = -25.715(12)  &-&92500 \times (\beta_c- 0.22165462) \nonumber \\ 
                   &-&1390  \times (\nu-0.63002)  \nonumber \\
                   &-&0.244 \times (\omega-0.832)
\end{eqnarray}
and
\begin{eqnarray}
a_+ b_+ = 3.87(28) &-& 130 0000 \times (\beta_c- 0.22165462) \nonumber \\
                   &-& 2900 \times (\nu-0.63002) \nonumber \\
                   &+& 13 \times (\omega-0.832) \;.
\end{eqnarray}
The number given in $()$ is the variance over all choices of $J$, $K$, $L$ and
$M$ with $\chi^2/$d.o.f.$< 1.073$. It might serve as a lower bound of the 
systematic error of the analysis of the series. Since the estimates for 
$E_{ns}$ and $C_{ns}$ obtained here are in good agreement with those obtained 
from the finite size analysis of Monte Carlo data given above, we are confident
that also in the case of $a_+$ and $a_+ b_+$ the variance over the 
choices of $J$, $K$, $L$ and $M$ is a realistic estimate of the systematical 
error.  Analysing the series for the free energy density itself
we get 
\begin{equation}
-f_{ns} = \ln 2  + 0.0847028611(4) + 0.99 \times (\beta_c- 0.22165462)
                 + 0.000001 \times (\nu-0.63002)
\end{equation}
The estimate of $f_{ns}$ strongly depends on the input value for $\beta_c$. 
The dependence on $\nu$ is small and that on $\omega$ can be ignored.

In order to calculate the energy density that is needed as input to compute
the thermodynamic Casimir force we picked, to some extend ad hoc, 
the approximant characterised by $J=7$, $K=7$, $L=5$ and $M=6$ which is 
characterized by the fact that the order of all four polynomials is similar, 
$\chi^2/$d.o.f$=1.029$ and $E_{ns}=0.9906063$ for $\beta_c=0.22165462$, 
$\nu=0.63002$ and $\omega=0.832$ fixed.  Comparing with other acceptable 
choices for  $J$, $K$, $L$ and $M$  we find that e.g. for $\beta=0.2216$ the 
differences are of the order $10^{-7}$ and for $\beta=0.22$ of the order
$10^{-8}$. Compared with the statistical error of 
$[E(L_0+d/2,\beta)-E(L_0-d/2,\beta)]/d$, see eq.~(\ref{DeltaE}), errors
of this size are negligible.

In the low temperature phase,  Vohwinkel \cite{Vohwinkel} computed
the energy density as a series in $u=\exp(-4 \beta)$ up to $O(u^{32})$. 
Unfortunately in this case there is no choice of $J$, $K$, $L$ and $M$ that 
allows to fit our Monte Carlo data down to $\beta=0.2219$.  The best 
that we could find are the two choices  $J=9$, $K=6$, $L=7$ and $M=13$ and
$J=20$, $K=6$, $L=3$  and $M=6$ that fit our Monte Carlo data with
an acceptable $\chi^2/$d.o.f. for $\beta \ge 0.228$ and  $\beta \ge 0.231$,
respectively. The linear 
combination $0.8155 E_{9,6,7,13} + 0.1845 E_{20,6,3,6}$  fits all of 
our data in the low temperature phase with $\chi^2/$d.o.f.$=1.25$. 

Since this result is not fully satisfying, we fitted our data with 
various ansaetze based on eq.~(\ref{Ebehave}). In particular 
the ansatz 
\begin{equation}
\label{finalE}
 E = E_{ns} - C_{ns} t + d_{ns} t^2 
   + a_{-} (-t)^{1-\alpha} + a_{-} b_{-} (-t)^{1-\alpha+\theta}  
   + b (-t)^{2-\alpha} + c (-t)^{2-\alpha-\theta}
\end{equation}
fits our data up to $\beta=0.246$ with $\chi^2/$d.o.f.$=1.15$, where we 
fixed $E_{ns}=0.9906065$, $C_{ns}=-29.07$, $\alpha=0.10994$ and $\omega=0.832$. 
Fitting all 55 data points up to $\beta=0.241$ we get for the free parameters
$a_{-}=47.9436$, $a_{-} b_{-} = -16.336$, $b=-363.5$, $d_{ns}=269.2$ and 
$c=287.3$.  
In order to calculate the bulk energy that is needed for the computation of the
thermodynamic Casimir force we used for $\beta \ge 0.228$ 
the linear combination $0.8155 E_{9,6,7,13} + 0.1845 E_{20,6,3,6}$ of 
approximants and for $0.228 > \beta \ge \beta_c$ we used
eq.~(\ref{finalE}) together with the results for the free parameters 
quoted above. For a quite large range of $\beta$ the two approaches to 
represent the bulk energy give consistent results. For 
$0.2219 \le \beta \le 0.2394$ the difference between the two is less than 
$3 \times 10^{-6}$. The deviation of our result from that of ref. \cite{Feng}
is typically of the order $10^{-5}$.  

Taking into account various fits and in particular computing the dependence
of the result on the values of the input parameters, we arrive at
\begin{eqnarray}
 a_{-} = 47.96(1) &+& 2350000 \times (\beta_c - 0.22165462) \nonumber \\
                  &+& 2500 \times (\nu-0.63002) \nonumber \\
                  &-& 0.16 \times (\omega-0.832) \nonumber \\
                  &-& 0.44 \times (C_{ns}-29.1) \nonumber \\
                  &-& 3700  \times (E_{ns}-0.9906065)
\end{eqnarray}
and hence 
\begin{equation}
 \frac{A_+}{A_-} = - \frac{a_+}{a_-} =  0.5362(20) 
\end{equation}
which is fully  consistent with the estimate 
$A_+/A_-=0.536(2)$ obtained by studying the Blume-Capel model 
at $D=0.655$ \cite{myamplitude}.
Note that the error of our estimate of $A_+/A_-$ is dominated by the 
uncertainty of $C_{ns}$ that we use as input for our fits in the low 
temperature phase. Here we took the error of the estimate obtained 
from the finite size scaling analysis at the critical point, 
eq.~(\ref{Cnonsingular}). The systematic error of the estimate obtained 
from the analysis of the high temperature series is likely smaller, but
difficult to estimate.
The authors of \cite{BuPe11} quote $A_+/A_-=0.530(3)$ which is slightly 
smaller than our results. For a summary of estimates presented in the 
literature see table IV or ref. \cite{BuPe11}.


\begin{thebibliography}{99}

\bibitem{WiKo}
K. G. Wilson and J. Kogut,
Phys.\ Rep.\ C {\bf 12}, 75 (1974).

\bibitem{Fisher74}
M. E. Fisher,
Rev.\ Mod.\ Phys.\ {\bf 46}, 597 (1974).

\bibitem{Fisher98}
M. E. Fisher,
Rev.\ Mod.\ Phys.\ {\bf 70}, 653 (1998).

\bibitem{PeVi02}
A. Pelissetto and E. Vicari, [cond-mat/0012164],
Phys.\ Rept.\ {\bf 368}, 549 (2002).

\bibitem{wegner71} F. J. Wegner, J. Math. Phys. {\bf 10}, 2259 (1971).

\bibitem{Wegner-76}
F. J. Wegner, in
{\em Phase Transitions and Critical Phenomena},
edited by C.~Domb and M.~S.~Green
(Academic Press, New York, 1976), Vol.\ 6.

\bibitem{mycritical}
M. Hasenbusch, 
[arXiv:1004.4486],
Phys. Rev. B {\bf 82}, 174433 (2010) 

\bibitem{NewmanRiedel}
K. E. Newman and E. K. Riedel, Phys.\ Rev.\ B {\bf 30}, 6615 (1984).

\bibitem{pisa97}
M. Campostrini, A. Pelissetto, P. Rossi and E. Vicari,
[cond-mat/9705086],
Phys.\ Rev.\ E {\bf 57}, 184 (1998). 

\bibitem{Barber}
M. N. Barber, ``Finite-size Scaling''
in {\sl Phase Transitions and Critical Phenomena, Vol. 8,}
eds. C. Domb and J. L. Lebowitz, (Academic Press, 1983)

\bibitem{BinderS}
K. Binder, ``Critical Behaviour at Surfaces''
in {\sl Phase Transitions and Critical Phenomena, Vol. 8,}   
eds. C. Domb and J. L. Lebowitz, (Academic Press, 1983)

\bibitem{Diehl86}
H. W. Diehl, {Field-theoretical Approach to Critical Behaviour at Surfaces} in
{\sl Phase Transitions and Critical Phenomena},
edited by C. Domb and J.L. Lebowitz, Vol. 10 (Academic, London 1986) p. 76.

\bibitem{Diehl97}
H. W. Diehl,
[cond-mat/9610143], 
Int.\ J.\ Mod.\ Phys.\ B {\bf 11}, 3503 (1997).

\bibitem{FiGe78}
M. E. Fisher and P.-G. de Gennes,
CR\ Seances Acad.\ Sci.\, Ser.\  B {\bf 287},  207 (1978).

\bibitem{Krech}
M. Krech, {\sl The Casimir Effect in Critical Systems}
(World Scientific, Singapore, 1994)

\bibitem{Dantchev1}
Daniel Dantchev, Michael Krech, and S. Dietrich, 
[cond-mat/0305596], Phys. Rev. E {\bf 67}, 066120 (2003).

\bibitem{Dantchev2}
Daniel Dantchev, Frank Schlesener,  and S. Dietrich, 
[cond-mat/0703122], 
Phys. Rev. E {\bf 76}, 011121 (2007).        

\bibitem{GaCh99}
R. Garcia and M. H. W. Chan,
Phys.\ Rev.\ Lett. {\bf 83},  1187  (1999).

\bibitem{GaScGaCh06}
A. Ganshin, S. Scheidemantel, R. Garcia, and M. H. W. Chan,
[cond-mat/0605663],
Phys.\ Rev.\ Lett. {\bf 97}, 075301 (2006).

\bibitem{FuYaPe05}
M. Fukuto, Y. F. Yano and P. S. Pershan, Phys.\ Rev.\
Lett.\ {\bf 94}, 135702 (2005).

\bibitem{Nature}
C. Hertlein, L. Helden, A. Gambassi, S. Dietrich, and C. Bechinger,
Nature (London) 451, 172 (2008).

\bibitem{GaMaHeNeHeBe09}
A. Gambassi, A. Macio\l ek, C. Hertlein, U. Nellen, L. Helden, C. Bechinger,
and S. Dietrich,
[arXiv:0908.1795],
Phys.\ Rev.\ {\bf E} 80, 061143 (2009).

\bibitem{Hucht}
A. Hucht, [arXiv:0706.3458], 
Phys.\ Rev.\ Lett.\ {\bf 99}, 185301 (2007). 

\bibitem{VaGaMaDi08}
O. Vasilyev, A. Gambassi, A. Macio\l ek, and S. Dietrich,
[arXiv:0812.0750], 
Phys.\ Rev.\ E {\bf 79}, 041142  (2009).

\bibitem{VaGaMaDi07}
O. Vasilyev, A. Gambassi, A. Macio\l ek, and S. Dietrich,
[arXiv:0708.2902],
Europhys.\ Lett.\ {\bf 80}, 60009 (2007).

\bibitem{Ga09}
A. Gambassi, [arXiv:0812.0935],
J.\ Phys.\ Conf.\ Series {\bf 161},  012037  (2009).

\bibitem{myXY}
M. Hasenbusch,
[arXiv:0905.2096], J. Stat. Mech. (2009) P07031 

\bibitem{mybreaking}
M. Hasenbusch, 
[arXiv:1005.4749], 
Phys.\ Rev.\  B {\bf 82}, 104425 (2010). 

\bibitem{ourXY}
M. Campostrini, M. Hasenbusch, A. Pelissetto, and E. Vicari,
[cond-mat/0605083], 
Phys. Rev. B {\bf 74}, 144506 (2006)

\bibitem{VaMaDi11}
O. Vasilyev, A. Macio\l ek,  and S. Dietrich,
[arXiv:1106.5140],
Phys.\ Rev.\ E {\bf 84}, 041605 (2011).

\bibitem{mycrossing}
M. Hasenbusch,
[arXiv:1012.4986], Phys. Rev. B {\bf 83}, 134425 (2011)

\bibitem{DeBl04}
Y. Deng and H. W. J. Bl\"ote,
Phys.\ Rev.\ E {\bf 70}, 046111 (2004).

\bibitem{myamplitude}
M. Hasenbusch, 
[arXiv:1004.4983],
Phys.\ Rev.\ B {\bf 82}, 174434 (2010).

\bibitem{pisaseries}
M. Campostrini, A. Pelissetto, P. Rossi and E. Vicari,
[cond-mat/0201180],
Phys.\ Rev.\ E {\bf 65}, 066127  (2002). 

\bibitem{HaPi93}
M. Hasenbusch and K. Pinn, 
[hep-lat/9310013], Physica A {\bf 203}, 189 (1994).

\bibitem{Wolff}
U. Wolff,
Phys.\ Rev.\ Lett.\ {\bf 62}, 361 (1989).

\bibitem{BrTa89}
R. C. Brower and P. Tamayo,
Phys.\ Rev.\ Lett.\ {\bf 62}, 1087 (1989).

\bibitem{twister}
M. Saito and M. Matsumoto,
``SIMD-oriented Fast Mersenne Twister: 
a 128-bit Pseudorandom Number Generator'', 
in
{\sl Monte Carlo and Quasi-Monte Carlo Methods 2006},
edited by A. Keller, S. Heinrich, H. Niederreiter, (Springer, 2008);
M. Saito, Masters thesis, Math. Dept., Graduate School of science,
Hiroshima University, 2007.
The source code of the program is provided at
``http://www.math.sci.hiroshima-u.ac.jp/$\sim$m-mat/MT/SFMT/index.html''

\bibitem{MH93A}
M. Hasenbusch,
[hep-lat/9209016],
J. Phys. I (France) {\bf 3}, 753 (1993).

\bibitem{MH93B}
M. Hasenbusch, 
Physica A {\bf 197}, 423  (1993).

\bibitem{ArFu03} H. Arisue and T. Fujiwara, Phys. Rev. E {\bf 67}, 066109
(2003), there is a typo in the 42th order term, the correct
value appears in hep-lat/0209002.

\bibitem{Vohwinkel}
C. Vohwinkel, Phys. Lett. B {\bf 301}, 208 (1993); and private communication.


\bibitem{HuGrSc11}
Alfred Hucht, Daniel Gr\"uneberg, Felix M. Schmidt,
[arXiv:1012.4399], 
Phys. Rev. E {\bf 83}, 051101 (2011).

\bibitem{wall}
M. Hasenbusch, K. Pinn and S. Vinti, [hep-lat/9806012],
Phys.\ Rev.\ B {\bf 59}, 11471 (1999).

\bibitem{KaRiMe01}
J. Kaupu$\breve{\mbox{z}}$s, J. Rim$\breve{\mbox{s}}$$\bar{\mbox{a}}$ns, 
and R. V. N. Melnik,
[arXiv:1103.0469],
Ukr. J. Phys. {\bf 56}, 845 (2011).

\bibitem{DeBl03X}
Y. Deng and H. W. J. Bl\"ote,
Phys.\ Rev.\ E {\bf 68}, 036125 (2003)

\bibitem{BuPe11}
P. Butera and M. Pernici, [arXiv:1012.5004],
Phys.\ Rev.\  B {\bf 83}, 054433 (2011) 

\bibitem{Feng}
X. Feng and H. W. J. Bl\"ote, [arXiv:0912.1467], 
Phys. Rev. E {\bf 81}, 031103 (2010).

\bibitem{BhCHLW94}
G. Bhanot, M. Creutz, I. Horvath, J. Lacki, and J. Weckel,            
Phys. Rev. E {\bf 49}, 2445 (1994).

\bibitem{Guttmann}
A. J. Guttmann, {Asymptotic Analysis of Power-Series Expansions} in
{\sl Phase Transitions and Critical Phenomena},
edited by C. Domb and J.L. Lebowitz, Vol. 13 (Academic, London 1989) p. 71.





%



\end{thebibliography}
\end{document}